\documentclass[aps,jap,twocolumn,amsmath,amssymb]{revtex4}

\usepackage{epsfig}
\graphicspath{{./}}
\newcommand{\etal}{\textit{et al.}\:}

\begin{document}

\title{Mechanism for Spontaneous Growth of Nanopillar Arrays \\
in Ultrathin Films Subject to a Thermal Gradient}
\author{Mathias~Dietzel and Sandra M.~Troian\footnote{Corresponding author: stroian@caltech.edu}}
\affiliation{California Institute of Technology\\
T. J. Watson Sr. Laboratories of Applied Physics\\
1200 E. California Blvd.\\
MC 128-95, Pasadena, CA 91125}
\date{\today}

\begin{abstract}
Several groups have reported spontaneous formation of
periodic pillar-like arrays in molten polymer nanofilms confined within
closely spaced substrates maintained at different temperatures. These formations
have been attributed to a radiation pressure instability caused by acoustic phonons.
In this work, we demonstrate how variations in the thermocapillary stress along the nanofilm interface
can produce significant periodic protrusions in any viscous film
no matter how small the initial transverse thermal gradient.
The linear stability analysis of the interface evolution equation
explores an extreme limit of B\'{e}nard-Marangoni flow peculiar
to films of nanoscale dimensions in which hydrostatic forces are
altogether absent and deformation amplitudes are small in comparison to the
pillar spacing. Finite element simulations of the full nonlinear equation
are also used to examine the array pitch and growth rates beyond the linear regime.
Inspection of the Lyapunov free energy as a function of time
confirms that in contrast to typical cellular instabilities in
macroscopically thick films,
pillar-like elongations are energetically preferred in nanofilms. Provided there occurs no dewetting
during film deformation, it is shown that fluid elongations continue to grow
until contact with the cooler substrate is achieved. Identification of
the mechanism responsible for this phenomenon may facilitate fabrication of
extended arrays for nanoscale optical, photonic and biological applications.
\end{abstract}


\maketitle

\section{Introduction}

The manufacture of ultra small optical and electronic components is nowadays
based on optical lithography techniques whereby a geometric pattern defined by
a photomask is transferred onto a photosensitive resist layer by exposure to UV
light. Various chemical treatments are then used to embed the positive or
negative image of this pattern onto a material film beneath the photoresist.
While this commercial technique can generate feature sizes below
100 nm, there are certain disadvantages inherent in the patterning process
\cite{WallraffHinsberg:ChemRev1999}. For example, multiple step-and-repeat
processes are required for deposition, exposure and removal of the photoresist
layers for constructing three dimensional components.
Inhomogeneities in the photoresist layer thickness,
composition, exposure dose or developer concentration can cause significant surface
roughness and scattering losses which diminish performance of optical or electronic
components. Optical lithography is also inherently a
two-dimensional technique whereby three dimensional components are fabricated
layer upon layer. The process requires that the supporting
substrates be rigid and flat, posing challenges for the fabrication of curved
or complex shaped components. In an effort to eliminate such constraints
while reducing fabrication time and cost, researchers have been exploring
alternative, lower resolution patterning techniques such as ink-jetting
\cite{FullerJacobson:JMEMS2002}, gravure printing \cite{MillerTroian:APL2003},
direct-write \cite{GratsonLewis:Nature2004}, micro-moulding
\cite{HeckeleSchomburg:JMicroMechEng2004} and nanoimprinting
\cite{ChouRenstrom:APL1995,Guo:AdvMat2007,Rogers:ChemRev2007}. These
methods are more adaptable to new materials and pattern
layouts; however, multiple etching steps are still required and
device performance is still not comparable to those fabricated by conventional means.
The materials of choice tend to be inks,
colloidal suspensions and polymer melts \cite{DelCampoArzt:ChemRev2008}, which
are not only less costly but whose composition can be tuned to optimize functionality.

Some groups have been investigating less conventional means of film patterning by
exploiting the self-assembling character of structures formed by hydrodynamic instabilities
in thin films. Examples include
dewetting induced by chemically templated substrates \cite{Han:Polymer2003},
capillary breakup on rippled substrates \cite{Mayr:JAP2008}, island formation in
ferroelectric oxide films \cite{Alexe:APL2003},
elastic contact instabilities in hydrogels \cite{Subramanian:Macromol2006} and evaporative
instabilities in metal precursor suspensions \cite{Cuk:APL2000}.
The use of fluid instabilities for controlled formation of large area, periodic
arrays provides an interesting approach for future development of non-contact, resistless
lithography.

It is well known that liquid films with dimensions in the micron to
nanometer range manifest exceedingly large surface to volume ratios. As
such, small liquid structures can respond instantaneously to external modulation of surface
forces. This sensitivity to surface manipulation has been successfully used
to control the motion of small liquid volumes for micro-, bio- and optofluidic
applications \cite{DarhuberTroian:ARFM2005}. For example,
tangential stresses based on thermocapillary forces
have been used to steer
\cite{DarhuberWagner:APL2003,DarhuberReisner:POF2003,DarhuberWagner:JMEMS2003},
mix \cite{DarhuberTroian:RoySoc2004} and shape
\cite{DietzelPoulikakos:POF2005} thin films and droplets on demand. Since the surface
tension of liquids varies with temperature, thermal distributions
can be applied directly to a supporting substrate to generate lateral gradients
which drive the flow of liquid
toward selected regions of a substrate.
In this work, we examine systems comprising of liquid nanofilms subject to a transverse
temperature gradient which have been observed to produce nanopillar
arrays which grow and elongate in the direction of a cooler target substrate.
The spontaneous formation of 3D large area arrays offers exciting possibilities for non-contact, resistless,
one step fabrication of optical and photonic structures.
Since solidification of the emergent molten structures occurs \textit{in-situ} upon
removal of the thermal gradient, it is anticipated that the resulting nanostructures
will manifest specularly smooth interfaces, a distinct advantage for optical applications.

\subsection{Formation of nanopillar arrays in molten polymer nanofilms}

\begin{figure}[htp]
\includegraphics[width=7.5cm]{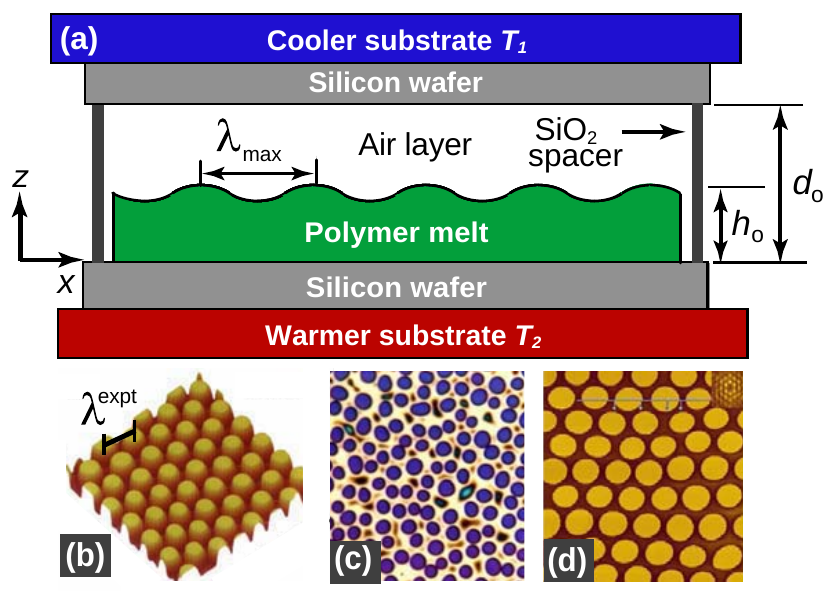}
\caption{(a) Sketch of experimental setup for formation of nanopillar
arrays. Initial thickness of flat nanofilm is denoted by $h_o$; gap spacing
in between silicon substrates is denoted by $d_o$. Length scale $\lambda_\textrm{max}$
represents theoretical prediction for pillar spacing; $\lambda^\textrm{expt}$ represents
experimentally measured values. (b) AFM image of PMMA pillars
\cite{ChouZhuang:JVacScieTech1999}: $h_o = 95\:\textrm{nm}$, $d_o =
260\:\textrm{nm}$, $\Delta T$ unknown, $\lambda^\textrm{expt}=3.4\:\mu \textrm{m}$.
(c) Optical micrograph of PS pillars
\cite{SchaefferSteiner:EurophysLett2002}: $h_o = 100\:\textrm{nm}$, $d_o =
285\:\textrm{nm}$, $\Delta T = 46\:^\textrm{o}\textrm{C}$,
$\lambda^\textrm{expt} = 2.9 \pm 0.6\:\mu$m. (d) AFM image of PMMA
pillars \cite{PengHan:Polymer2004}: $h_o = 100\:\textrm{nm}$, $d_o =
163\:\textrm{nm}$, $\Delta T = 10\:^\textrm{o}\textrm{C}$,
$\lambda^\textrm{expt} = 6.5\:\mu \textrm{m}$.} \label{Fig:setup}
\end{figure}

The typical experimental setup leading to spontaneous formation of
nanopillar arrays is shown in Fig.~\ref{Fig:setup}(a). Polymers such as polystyrene (PS) or
poly(metylmetacrylate) (PMMA) are first spun cast onto a clean, flat silicon wafer to
an initial thickness $h_o$ of the order of a few hundred nanometers. The coated
wafer is then overlay with a second silicon wafer containing vertical
spacers along the periphery to ensure an air gap above the polymer film. The wafer separation
distance, $d_o$, is normally several hundred nanometers. The bottom and
top wafers are maintained at different temperatures above the polymer glass transition
temperature to ensure a flowing liquid film. In all the experiments reported in the literature,
$\Delta T = T_2 - T_1 \approx 10 - 50 ^\textrm{o}
\textrm{C}$. Next, we review the experimental results of three
independent groups reporting observations and measurements of nanopillars arrays.

\subsubsection{Experiments by Chou \etal}

Chou \etal \cite{ChouZhuang:JVacScieTech1999,ChouGuo:APL1999} appear to have
been the first group to report nanopillar formation
in ultrathin polymer films. In their experiments, they studied
low molecular weight PMMA (approx 2K), which was first spun cast to a film thickness
of 100 nm onto a cleaned silicon wafer and then
annealed at $80\:^\textrm{o} \textrm{C}$ to drive off residual solvent. The annealed film was
then placed within the assembly shown in Fig.~\ref{Fig:setup}(a),
where the top wafer had been treated with a nonstick
coating to prevent polymer attachment after
solidification. The underside of the top wafer was either flat or patterned with a rectangular relief
structure a few tens of microns in width
and about $0.3 \mu$m tall. In all experiments
reported, there was no imposed temperature difference
between the top and bottom wafers ($T_2 - T_1 = 0$). Instead, the entire assembly was cyclically
heated from room temperature to either $130\:^\textrm{o} \textrm{C}$ or
$170\:^\textrm{o} \textrm{C}$, well above the polymer glass transition
temperature $T_\textrm{g} = 103\:^\textrm{o}\textrm{C}$  \cite{Mark:AIPPress1996}
to ensure a softened film. The heating cycle persisted for 5-80 minutes
with no noticeable difference in pattern formation if the air gap was
replaced by a vacuum at 0.3 Torr. In cases where the PMMA coated
wafer was not overlay by a top wafer and simply exposed to open air, no
protrusions were observed to form. When the top wafer was placed in close
proximity to the melt surface i.e. $(d_o - h_o) \approx 165\:\textrm{nm}$ , nanopillar
arrays with in-plane hexagonal symmetry were obtained, as in the image shown in
Fig.~\ref{Fig:setup}(b). These elongations were measured to have a
diameter and pitch (i.e. pillary spacing) of a few microns; their overall height closely matched the
gap distance $d_o$ separating the two wafers. AFM images of the resulting structures
after solification revealed pillars with a flat top and fairly
straight sidewalls. Chou \etal attributed the formation of these elongations
to an image-charge induced electrohydrodynamic instability caused by
non-uniform distribution of charges on the relief surface.
Chou \etal also noted that thermal gradients might
be playing a role but that
Rayleigh-B\'{e}nard or B\'{e}nard-Marangoni cellular convection was unlikely since the
initial film thicknesses were far too small to overcome the relevant critical
numbers required for instability \cite{ChouGuo:APL1999}.

\subsubsection{Experiments and modeling efforts by Sch\"affer \etal}

Soon thereafter, Sch\"affer and co-workers
\cite{SchaefferSteiner:EurophysLett2002,SchaefferSteiner:Macromolecules2003,Schaeffer:UKonstanz2001}
used a similar setup as in Fig.~\ref{Fig:setup}(a) where the two
confining wafers were purposely set to different temperatures such that
$T_2 > T_1$. They first spun cast high molecular weight films of PS ($T_\textrm{g} = 95\:^\textrm{o} \textrm{C}$
\cite{Mark:AIPPress1996}, mol. wt. 108 kg/mol) dissolved in toluene onto a silicon wafer down
to an initial thickness $80\:\textrm{nm} \lesssim h_o
\lesssim 130\:\textrm{nm}$. It appears that these films were not annealed
to drive out residual solvent after spin casting,
which may have led to overestimates in the reported values of $h_o$ (discussed further in Section III).
The wafer separation distance
ranged from $100\:\textrm{nm} \lesssim d_o
\lesssim 600\:\textrm{nm}$. The bottom wafer was then heated to $T_2 =
170\:^\textrm{o} \textrm{C}$; the top wafer was cooled to a temperature
above $T_\textrm{g}$ such that $\Delta T=
T_2 - T_1$ ranged from $10 \leq \Delta T \leq 55\:^\textrm{o}
\textrm{C}$. The small wafer separation distances give rise to very large
transverse thermal gradients of the order of $\Delta T/d_o \sim 10^6 -
10^8\:^\textrm{o} \textrm{C/cm}$. After subjecting the PS film to the
thermal gradient overnight, the sample was quenched to room temperature
and the top wafer removed. As in Chou \etal, the top
wafer had been treated with a silanized monolayer in order to prevent adhesion
of the PS. After solidification and removal of the top wafer, the films were observed to
contain periodic nanopillar arrays, as shown in Fig.~\ref{Fig:setup}(c).
To study the influence of the wafer separation distance $d_o$ on the pillar formation process,
Sch\"affer \etal used a tilted plate geometry in all their experiments
where the top wafer was inclined with respect
to the bottom one by about $1 \mu$m over a distance of
1 cm, corresponding to an inclination angle of about $0.0057 ^\textrm{o}$.
This modification allowed simultaneous measurement
of the array pitch as a function of $d_o$ within a single run.
Sch\"affer \etal conducted a comprehensive set of experiments and determined
the influence of the initial film thickness $h_o$,
the wafer separation distance $d_o$, and temperature drop $\Delta T$ on the
pillar separation distance $\lambda$. They ruled out any
electrostatic effects by purposely grounding the confining wafers.

As noted both by Chou \etal \cite{ChouZhuang:JVacScieTech1999} and Sch\"affer \etal
\cite{SchaefferSteiner:Macromolecules2003}, films ranging in thickness from millimeters
to centimeters subject to a transverse thermal gradient are known to
develop cellular instabilities which lead to periodic surface deflections at the air/liquid interface
due either to Rayleigh-B\'{e}nard (RB) or B\'{e}nard-Marangoni (BM) convection
\cite{Probstein:Wiley1994}. These instabilities, however, generate very shallow
corrugations and not pillar-like protrusions as observed
in nanofilms. Onset of instability requires that the critical Rayleigh number
$Ra_\textrm{onset}$ for buoyancy driven flow (which scales as $h^4_o$) or the
critical Marangoni number $Ma_\textrm{onset}$ for thermocapillary flow (which
scales as $h^2_o$) exceed 660 - 1700 or 50-80, respectively, depending on the
boundary conditions. In the nanofilm experiments,
the corresponding values are estimated to be $Ra
\approx 10^{-16}$ and $Ma \approx 10^{-8}$, orders of magnitude less than
required for onset of instability.

Sch\"affer \etal therefore proposed a different mechanism for instability and interfacial
deformation based on a novel radiation pressure model. They hypothesized that
low frequency acoustic phonons (AP) can reflect coherently
from the interfaces of the molten film over distances of the order of the film thickness
despite that the melt is in an amorphous state.
These low frequency modes are postulated to generate a significant destabilizing
radiation pressure while conducting little heat. By contrast, the high frequency modes
are expected to propagate diffusively with little interfacial resistance and
therefore little interfacial pressure.
These modes, however, are essential for establishing the steady-state heat flux across the
air and melt layers. The mechanism described represents a kind of
acoustic analogue of the radiation pressure caused by optical phonon reflections in closely
spaced metal plates placed in vacuum, known to generate the Casimir
interaction force \cite{MorariuSteiner:PRL2004}. Since the air/melt interface is liquid-like
and therefore deformable, the acoustic phonons in the polymer melt
are believed to generate an outwardly oriented radiation pressure, which
counteracts the stabilizing force of surface tension;
infinitesimal surface deflections can therefore grow into sizeable protrusions. Sch\"affer
\etal developed a detailed hydrodynamic model based on the slender gap approximation
for describing the evolution equation for the film thickness, $h(x,y,t)$.
A linear stability analysis of this evolution equation leads to an
analytic expression for the wavelength corresponding to the fastest growing unstable mode, namely
\begin{equation}
\lambda^\textrm{AP}_{\textrm{max}} = 2 \pi h_o \sqrt{\frac{\gamma \: u_p}{Q
(1-\kappa) k_{\textrm{air}} \Delta T}}\left[\frac{d_o}{h_o} +\kappa -
1\right]~,
\label{Eq:APwavelength}
\end{equation}
where $\gamma$ denotes the surface tension of the polymer melt, $u_p$ is the
speed of sound in the polymer melt and $\kappa =
k_{\textrm{air}}/k_{\textrm{melt}}$ denotes the ratio of thermal
conductivity of air to that of the polymer melt. The superscript AP differentiates
this expression from the one to be derived for a thermocapillary model (TC).
The material constants in Eq. (\ref{Eq:APwavelength}) are
evaluated at the substrate temperature $T_2$. The parameter $Q$ represents
the acoustic quality factor determined from the phonon reflection and
transmission coefficients corresponding to the four media constituting the
system, namely the bottom silicon wafer, the polymer melt, the overlying air
layer and the top silicon wafer. Positive values of $Q$
lead to film destabilization and the formation of nanopillar arrays.
Sch\"affer \etal compared the prediction for $\lambda^\textrm{AP}_{\textrm{max}}$
directly with the pillar spacings obtained in experiment, $\lambda^\textrm{expt}$. A least squared fit of the
experimental data to the model with $Q$ and $u_p$ as fitting parameters produced good agreement
(see dashed curves in Fig.\ref{Fig:comp_exptheo}(b) ).
In particular, it was shown that the value of $Q$ did not vary with
$h_o$, $d_o$ or $\Delta T$.
The acoustic quality factor $Q$ seemed to depend on the choice of substrate;
$Q = 6.2$ was obtained for the silicon/air/PS/silicon system, while $Q = 83$
for films supported by a silicon wafer coated with a 100 nm layer of gold.
Unfortunately, it was reported \cite{Schaeffer:UKonstanz2001} that the measurements of
$\lambda^\textrm{expt}$ included not only pillar formations but lamellar structures, spirals
and other periodic formations caused either by defects in the initial film or by
prolonged contact with the cooler substrate. In many cases, protrusions had undergone
reorganization while in contact with the cooler wafer. In addition, measurements of pattern periodicity
were obtained long after contact with the upper substrate and subsequent solidification. Comparison of these
measurements to a model based on linear instability is therefore problematic.

Sch\"affer \etal concluded that they had uncovered a novel instability in
nanofilms induced by a radiation pressure from interfacial reflections of
low frequency acoustic phonons. They noted that the frequency dependence for
propagation of acoustic phonons with large mean free path is highly unlikely
in low molecular weight polymers and that the instability would not be observed in
such systems since they lack the necessary glassy rheological response
\cite{SchaefferSteiner:Macromolecules2003}.
In a separate study, Sch\"affer \etal\cite{SchaefferSteiner:AdvMat2003}
also conducted experiments with relief structures patterned with complex patterns
held in close proximity to the polymer melt interface. The smallest values of $d_o/h_o$ lead
to well defined replicas in the polymer film.

\subsubsection{Experiments by Peng \etal}

Shortly following the work of Sch\"affer \etal,
Peng and co-workers \cite{PengHan:Polymer2004} used a
similar assembly as in Fig.~\ref{Fig:setup}(a) to study PMMA films with $h_o
\approx 100$ nm, $T_2=160\:^\textrm{o} \textrm{C}$, $130\:^\textrm{o}
\textrm{C} \leq T_1 \leq 150\:^\textrm{o} \textrm{C}$ and
$110 \leq d_o \leq 210\:\textrm{nm}$. They were able to obtain nanopillar
arrays after about $0.5 - 2.5\:\textrm{hrs}$; however, they did not
conduct a parametric study nor compare their measurements of the pillar
spacing with the prediction of Sch\"affer \etal. Fourier transforms of the
nanopillar arrays showed well defined hexagonal symmetry in some cases,
as shown in Fig.~\ref{Fig:setup}(d). In other experiments, the pillar formations adopted
either stripe or spiral symmetry. Peng \etal used a simple energy minimization argument
first introduced by Sch\"affer \etal to show that pattern selection between
stripe and hexagonal arrangements is merely controlled by the thickness of the
overlying air film, while spiral formations are likely caused by point defects in the film.
In a final experiment, Peng and co-workers successfully transferred nanopillar patterns
first formed in PMMA onto
an elastomeric film of poly(dimethylsiloxane) (PDMS) i.e. negative replication
of the original pattern. This demonstration outlined the ease with which
potential patterns can be transferred into subsequent films for applications involving
large area patterning.

\subsection{Motivation for this study}

In recent work \cite{DietzelTroian:PRL2009}, we re-examined the prevailing hypothesis
for pillar formation in nanofilms based on coherent reflections of acoustic phonons
in molten polymer nanofilms \cite{SchaefferSteiner:Macromolecules2003,SchaefferSteiner:AdvMat2003}.
Such a mechanism requires coherent phonon
propagation of the order of the film thickness in an amorphous fluid
layer. A review of the literature has shown that
acoustic phonon mean free paths of the order of 10-100 nm have only been
measured in \textit{solid} polymer nanofilms at frequencies of order 100 GHz
and at temperatures $-193 ^\textrm{o} \textrm{C} \leq T \leq 27
^\textrm{o} \textrm{C}$ \cite{MorathMaris:PhysRevB1996},
far below the temperatures used in the experiments described above.
Such long attenuation lengths,
however, are highly unlikely in molten amorphous films far above $T_\textrm{g}$
because of the degree of disorder present and the enhanced mobility of polymer chains at
temperatures above $T_g$.

Given that the free surface of thin liquid films is easily deformed by surface
stresses \cite{DarhuberTroian:ARFM2005}, we instead demonstrate in this work
that nanopillar formations are caused by the nanoscale analogue of the
long-wavelength B\'enard-Marangoni instability
\cite{ScrivenSterling:JFM1964,Smith:JFM1966,OronDavis:RevModPhys1997,VanHook:PRL1995},
previously investigated for film thicknesses ranging from several hundred
microns ($70 \lesssim h_o
\lesssim 270\:\mu \textrm{m}$ \cite{VanHook:JFM1997,VanHook:PRL1995}) to millimeters.
In macroscopically thicker films, film protrusions caused by thermocapillary flow
are stabilized by capillary and gravitational forces, such that only gentle surface deflections
are possible \cite{VanHook:JFM1997}. Onset of instability in such films requires that the inverse dynamic
Bond number $D^{\textrm{dyn}}_\textrm{onset}=\gamma_T \Delta T_\textrm{film} /\rho g {h_o}^2
\geq 2/3(1+F)^{-1}$, where $\rho$ is the liquid density, $\gamma_T \equiv
|d\gamma/dT|$, $\gamma$ is the liquid surface tension, $\Delta T_\textrm{film}$
is the temperature drop across the liquid layer, $F=(1-\kappa)/(D + \kappa -1)$
is an order one constant, $D=d_o/h_o$, and
$\kappa=k_\textrm{air}/k_\textrm{melt}$. Estimates corresponding to
the experiments of Sch\"affer \etal and Peng \etal indicate that
$D^{\textrm{dyn}} \gtrsim \textsf{O}(10^7)$ and $G \sim \textsf{O}(10^{-14})$.
These critical values lie
far beyond the regime previously investigated by Vanhook and co-workers
\cite{VanHook:JFM1997,VanHook:PRL1995} in which $D^{\textrm{dyn}}_\textrm{onset} \sim
\textsf{O}(10^{-1} - 1)$ and $G \sim \textsf{O}(10^{-1} - 10^2)$. These
estimates indicate that nanofilms dominated by thermocapillary flow should
always undergo instability. In what follows, we therefore propose an alternative mechanism to
the acoustic phonon model to help explain the formation of elongated structures
in liquid nanofilms subject to a transverse thermal gradient.
The analysis presented here
indicates that the experiments conducted by
Sch\"affer \etal and Peng \etal provide a rare window into the dynamics of the
less common long-wavelength B\'enard-Marangoni (BM) instability without interference from
the better known short-wavelength (BM) instability, which gives rise
to the beautiful cellular convection patterns often photographed.

There is an additional feature worth emphasizing in Fig.~\ref{Fig:setup}(a).
In the absence of a top wafer,
a transverse thermal gradient can still be established in a film heated from below
by natural or forced convection within the gas layer above the polymer melt.
Since the Biot number $\beta h_o/k_{\textrm{melt}}$ is
linearly proportional to the polymer film thickness $h_o$
(where $\beta$ is the heat transfer coefficient for natural convection), however,
this number will be small. As a result, the thermal gradient within
the viscous film will also be small and thermocapillary stresses at the interface may
be easily stabilized by capillary forces. This is probably the reason
why no fluid elongations were observed in the experiments of Chou \etal
in which the polymer melt was heated in open air. Use of a top substrate maintained
at a cooler temperature held in close proximity to the melt surface enforces a
sizeable transverse thermal gradient which can be used to
maximize and control thermocapillary flow.

In this work we demonstrate that the predominance of thermocapillary forces
along the free surface of molten nanofilms leads to a linearly unstable system
which forms periodic protrusions no matter how small the applied thermal gradient in any liquid
nanofilm, not just molten polymeric films.
The analysis corresponds to a limiting case of B\'{e}nard-Marangoni flow peculiar to viscous
films of nanoscale dimensions such that hydrostatic forces are
completely negligible and deformation amplitudes are small in comparison to the array pitch.
Predictions of the pillar spacing from the linear analysis as a function of the
substrate separation distance reveals good agreement with experiment.
Deviations are likely due to overestimates in the reported values of $h_o$ for unannealed
films, uncertainties in the measured values of $d_o$ caused by the use of a tilted upper plate,
and possible changes in wavelength caused by prolonged contact with the cooler substrate and
film solidification prior to measurements of the array pitch.
Finite element simulations of the full nonlinear equation
are also used to examine the array pitch and growth rates beyond the linear regime.
Inspection of the Lyapunov free energy as a function of time
confirms that in contrast to typical cellular instabilities in
macroscopically thick films,
pillar-like elongations are energetically preferred in nanofilms. Provided there occurs no dewetting
during film deformation, it is shown that fluid elongations continue to grow
until contact with the cooler substrate is achieved. Identification of
the mechanism responsible for this phenomenon may facilitate fabrication of
extended arrays for nanoscale optical, photonic and biological applications.

\section{Evolution of molten nanofilms subject to the slender gap approximation}

\subsection{Films confined by parallel substrates}

The molten layer is modeled as an incompressible Newtonian fluid
since the flow speeds and shear rates inherent in the
experiments described are very small. Consistent with the slender gap
approximation, all lateral dimensions are scaled by the pillar spacing distance
$L$, while all vertical scales are normalized by the initial film thickness
$h_o$ such that $(X,Y) = (x/L,y/L)$, $Z = z/h_o$, $H(X,Y,\tau) = h(x,y,t)/h_o$
and $D_o = d_o/h_o$. The pillar spacing $L$ will later be identified with the
wavelength of the maximally unstable mode, $\lambda_{\textrm{max}}$, obtained
from linear stability analysis. The conservation equations for mass and
momentum within the thin liquid film are given by
\begin{equation}
\label{Eq:mass} \partial U/\partial X + \partial V/\partial Y +
\partial W/\partial Z = 0
\end{equation}
\begin{equation}
\label{Eq:momentum_x} \epsilon Re \frac{DU}{D\tau} -\epsilon^2\left(
\frac{\partial^2 U}{{\partial X}^2} + \frac{\partial^2 U}{{\partial
Y}^2}\right) = -\frac{\partial P}{\partial X} + \frac{\partial^2 U}{{\partial
Z}^2}
\end{equation}
\begin{equation}
\label{Eq:momentum_y} \epsilon Re \frac{DV}{D\tau} -\epsilon^2\left(
\frac{\partial^2 V}{{\partial X}^2} + \frac{\partial^2 V}{{\partial
Y}^2}\right) = -\frac{\partial P}{\partial Y} + \frac{\partial^2 V}{{\partial
Z}^2}
\end{equation}
\begin{equation}
\label{Eq:momentum_z}
\epsilon^3 Re \frac{DW}{D\tau}
-\epsilon^2\left(\epsilon^2 \frac{\partial^2 W}{{\partial X}^2} +\epsilon^2
\frac{\partial^2 W}{{\partial Y}^2} + \frac{\partial^2 W}{{\partial
Z}^2}\right) = -\frac{\partial P}{\partial Z}.
\end{equation}
Equation (\ref{Eq:mass}) yields the scaling for the velocity components,
namely $\overrightarrow{U} = (U,V,W) =
(u/u_\textrm{c},v/u_\textrm{c},w/\epsilon u_\textrm{c})$, where $u_\textrm{c}$
represents the characteristic lateral speed set by thermocapillary flow. The
corresponding Reynolds number based on the initial film thickness is $Re = \rho
u_\textrm{c} h_o/\eta$, where $\rho$ and $\eta$ denote the polymer melt density
and viscosity. In what follows, the polymer viscosity is assumed
constant (i.e. a Newtonian fluid) and equal to $\eta = \eta(T_2)$ \cite{footnote1}.
The non-dimensional Lagrangian or substantial derivative is denoted by
$D/D\tau = \partial/\partial \tau + \overrightarrow{U}\cdot\nabla$ where $\tau
= u_\textrm{c} t/L$. The overall (dimensionless) pressure in the fluid is given by
\begin{equation}
P= \epsilon h_o (p + \phi)/(\eta u_\textrm{c})
\end{equation}
where $p$ is the (dimensional) capillary pressure and
$\phi$ represents contributions from hydrostatic pressure
(i.e. $\phi = g\:z$ where $g$ is the gravitational
constant) and disjoining pressure (e.g. van der Waals forces).

Within the slender gap approximation, $\epsilon^2=(h_o/L)^2\ll1$ and $\epsilon
Re \rightarrow 0$; all terms on the left hand side of Eqs.
(\ref{Eq:momentum_x}) - (\ref{Eq:momentum_z}) therefore vanish.  In this limit,
the pressure $P$ within the thin film is independent of the vertical coordinate $Z$.
Equations ~(\ref{Eq:momentum_x}) and (\ref{Eq:momentum_y}) can therefore be
integrated with respect to $Z$, subject to the boundary conditions (BCs) at the
liquid/solid and gas/liquid interface. Along the bottom substrate, it is assumed
that the melt obeys the no-slip condition i.e. $\overrightarrow{U_\parallel}=(U,V)=0$.
The dimensional stress jump across the air/melt interface \cite{Leal:CAMPress2007},
which accounts for both normal and tangential stresses, is given by
\begin{equation}
\label{Eq:totstressbalance}
(\textbf{T}_{\textrm{air}} - \textbf{T}_{\textrm{melt}}) \cdot \hat{n}
+ \nabla_s \gamma - \gamma \hat{n}(\nabla_s \cdot \hat{n}) = 0.
\end{equation}
Here, $\textbf{T}=- (p + \phi) \textbf{I} + 2 \eta \textbf{E}$ denotes the total
bulk stress tensor, where $\textbf{I}$ is the unit tensor and $\textbf{E}$ the rate of strain
tensor, $\hat{n}$ denotes the unit vector outwardly pointing from the melt interface,
$\nabla_s$ represents the surface gradient operator \cite{Deen:Oxford1998}
and $\gamma$ is the surface tension of the polymer melt in air.
Since the viscosity and density of air are negligible
in comparison to those of the melt, $\textbf{T}_{\textrm{air}}=0$.

Thermocapillary flow within the melt leads to a non-vanishing shear stress
$\nabla_s \gamma$ along the gas/liquid interface \cite{Leal:CAMPress2007}.
After a straightforward derivation, it can be shown within the slender gap
approximation \cite{OronDavis:RevModPhys1997} that the
tangential components of Eq. (\ref{Eq:totstressbalance}) reduce to
\begin{equation}
\label{Eq:tang_stress_x} {\partial U/\partial Z}|_{Z=H(X,Y,\tau)} =
\partial \Gamma/\partial X
\end{equation}
\begin{equation}
\label{Eq:tang_stress_y} {\partial V/\partial Z}|_{Z=H(X,Y,\tau)} =
\partial \Gamma/\partial Y
\end{equation}
where the surface gradient simplifies to $\nabla_s=\nabla_\parallel =
(\partial/\partial X, \partial/\partial Y)$.
The variable $\Gamma = \epsilon \gamma/(\eta u_\textrm{c})$ represents the
dimensionless surface tension. The gradients in surface tension
arise directly from thermal gradients along the melt interface i.e.
$\nabla_\parallel \gamma = (d \gamma/dT) \nabla_\parallel T$.
In dimensionless form, this relation is given by
\begin{equation}
\label{Eq:surftensgrad} \nabla_\parallel \Gamma = -\frac{\epsilon
\gamma_T}{\eta u_\textrm{c}} \nabla_\parallel T|_{Z=H} = -\overline{Ma}
\nabla_\parallel \Theta|_{Z=H}\,
\end{equation}
where $\Theta =(T-T_1)/(T_2-T_1)$, $\gamma_T = |d\gamma/dT|$, $\Delta T = T_2 -
T_1 > 0$, and the Marangoni number $\overline{Ma} = \epsilon \gamma_T \Delta
T/(\eta u_\textrm{c})$. In what follows, it is assumed that $T_2-T_1 > 0$; furthermore,
for the liquid films of interest, the surface tension decreases linearly with increasing
temperature $T$, which is reflected in the choice of the negative sign above.

The in-plane velocity components are therefore given by:
\begin{equation}
\label{Eq:velocities_lat}
\overrightarrow{U}_\parallel
=\left(\begin{array}{c}
U \\
V
\end{array}\right) = \left(\frac{Z^2}{2} -
H\:Z\right)\nabla_{\parallel}P + Z~ \nabla_{\parallel}\Gamma.
\end{equation}
Equation (\ref{Eq:velocities_lat}) represents a linear superposition
of pressure driven flow caused by variations in interfacial curvature and
hydrostatic forces, as
described by Eq. (\ref{Eq:normal_stress}), and
shear driven flow induced by thermocapillary stresses.
Substitution of Eq. (\ref{Eq:velocities_lat}) into Eq. (\ref{Eq:mass}) followed
by integration subject to the condition $W (X,Y, Z=0)=0$ gives the vertical
component of the velocity field,
\begin{equation}
\label{Eq:velocities_vert} W = \left(\frac{H Z^2}{2} -
\frac{Z^3}{6}\right){\nabla_{\parallel}}^2P +
\frac{Z}{2}\left(\nabla_{\parallel}P\cdot\nabla_{\parallel}H -
{\nabla_{\parallel}}^2\Gamma\right).
\end{equation}
The evolution equation for the moving interface can then be determined by
integration of Eq.~(\ref{Eq:mass}) from $0 \leq Z \leq H(X,Y,\tau)$ subject to
$W(X,Y,Z=0)=0$ and the kinematic boundary condition, $W|_{Z=H} = DH/D\tau =
\partial H/\partial \tau + \overrightarrow{U}|_{Z=H}\cdot\nabla_s H$.
The Leibnitz rule for differentiation gives
\begin{equation}
\label{Eq:tf_raw} \frac{\partial H}{\partial \tau} + \nabla_\parallel \cdot
\left(\int^{H(X,Y,\tau)}_0 \overrightarrow{U}_\parallel dZ\right) = 0.
\end{equation}
Substitution of Eq.(\ref{Eq:velocities_lat}) leads to the evolution equation for the
melt interface $H(X,Y,\tau)$, namely
\begin{equation}
\label{Eq:interface} \frac{\partial H}{\partial \tau} + \nabla_\parallel \cdot
\left(\frac{H^2}{2} \nabla_\parallel \Gamma - \frac{H^3}{3} \nabla_\parallel P
\right) = 0.
\end{equation}
It is expected that the slender gap approximation remains valid throughout the growth process
so long as $(d_o/L)^2 \ll 1$, which holds for all the experiments described.

Since the pressure in the film is independent of $Z$ to order $\epsilon^3 Re$, one can determine
its value by considering the normal stress balance at $Z=H$. The normal component of
Eq. (\ref{Eq:totstressbalance}) within the slender gap approximation
yields the total pressure in the film to order $\epsilon^2$:
\begin{equation}
\label{Eq:normal_stress} P = -\overline{Ca}\:^{-1} \nabla^2_\parallel H +
\overline{Ca}\:^{-1}\:\overline{Bo}\:H,
\end{equation}
where $\overline{Ca} = \eta u_\textrm{c}/(\gamma \epsilon^{3})$ and
$\overline{Bo} = \rho g L^2/\gamma$. Parameter estimates from the experiments of Sch\"affer \etal
indicate that $\overline{Ca}$ is of the order of $10^{1} - 10^{2}$ [using Eq. (\ref{Eq:u_c})]
while $\overline{Bo}$ is of the order of $10^{-5} - 10^{-6}$.
The hydrostatic contribution to the fluid pressure in Eq.~(\ref{Eq:normal_stress}) can therefore be
neglected altogether. The influence of disjoining pressure arising from van der Waals interactions
in films ranging from 10 - 100 nm in thickness \cite{OronDavis:RevModPhys1997} is also ignored in this work.
The flow induced by these molecular forces is weak in comparison to
flow induced by thermocapillary stresses, which are of considerable magnitude in the
experimental systems of interest. While disjoining pressure effects can be included in straightforward
fashion within $P$, they are not the primary mechanism for instability.
Furthermore, there is yet no consensus in the literature on the appropriate analytic form of the
disjoining pressure in cases where films are subject to large thermal gradients; most of the simplified
forms available in the literature are only appropriate for isothermal systems.
It is also assumed that any thermocapillary effects caused by solvent evaporation and subsequent
cooling of the interface \cite{BestehornThiele:EuroPhysJB2003} can be neglected. This assumption requires that
solvent evaporation be completed (either naturally or by film annealing)
before the film is inserted into the experimental assembly.

With these assumptions, the gradient of the Laplace pressure is
given by
\begin{equation}
\label{Eq:nstressgrad} \nabla_\parallel P = -\overline{Ca}\:^{-1}
\nabla^3_\parallel H - \epsilon^2 \nabla^2_\parallel H \nabla_\parallel \Gamma.
\end{equation}
The last term, which represents a correction to the Laplace pressure
due to local variation in surface tension, scales as $\epsilon^2$ and can be safely
ignored. The surface tension coefficient in the Laplace pressure \textit{only}
is therefore set to the value $\gamma=\gamma(T_2)$.

Determination of the interfacial stress conditions in Eqs. (\ref{Eq:tang_stress_x}) and (\ref{Eq:tang_stress_y})
requires knowledge of the thermal distribution
along $Z=H$, which can be obtained
from the energy equations \cite{OronDavis:RevModPhys1997} pertaining to the confined air/liquid bilayer
shown in Fig. \ref{Fig:setup}:
\begin{equation}
\label{Eq:energy} \epsilon Re Pr \frac{D\Theta}{D\tau}
-\epsilon^2\left(\frac{\partial^2 \Theta }{{\partial X}^2} + \frac{\partial^2
\Theta }{{\partial Y}^2}\right)= \frac{\partial^2 \Theta}{{\partial Z}^2}.
\end{equation}
Here, the Prandtl number $Pr = \nu/\alpha$ refers to the kinematic viscosity $\nu$ and
thermal diffusivity $\alpha$ of the corresponding air or liquid melt layer.
The Reynolds number $Re$, defined previously, is based on the corresponding layer
thicknesses. Despite that $Pr$ is of the order of $10^8 - 10^9$ for the polymer melts of interest,
the small gap approximation coupled with the vanishingly small value of $Re$
(see Tables \ref{Tbl:char_numbers} and \ref{Tbl:material_properties})
ensures that the left hand side of Eq.~(\ref{Eq:energy})
is completely negligible. In fact,
the slender gap approximation is well satisfied in all the experiments described earlier
since $\epsilon^2 \ll 1$, $\epsilon \, Re \ll 1$ and $\epsilon Re Pr \ll 1$.
The thermal analysis conveniently reduces to a one dimensional
thermal conduction problem for heat flow across an air/liquid bilayer
subject to isothermal boundary conditions at $Z=0$ and $Z=D_o$. The temperature distribution along the melt interface
is therefore given by, $\Theta|_{Z=H}
=(D_o-H)/[D_o+(\kappa - 1)H]$. Substitution of this solution into
Eq.~(\ref{Eq:surftensgrad}) yields
\begin{equation}
\label{Eq:surftensgrad_1DTD} \nabla_\parallel \Gamma = \frac{\kappa~
\overline{Ma} ~D_o \nabla_\parallel H}{[D_o+(\kappa-1)H]^2}.
\end{equation}

Substitution of Eqn.~(\ref{Eq:nstressgrad}) and (\ref{Eq:surftensgrad_1DTD})
into Eq.~(\ref{Eq:interface}) then yields the expression governing the motion of the
air/liquid interface, namely
\begin{equation}
\label{Eq:TCinterface}
\frac{\partial H}{\partial \tau}\!+\! \nabla_\parallel
\cdot \!\left[\frac{\kappa \, D_o \,\overline{Ma}\, H^2}{2\,[D_o + (\kappa -
1)H]^2}\nabla_\parallel H +\frac{H^3}{3\, \overline{Ca}} \nabla^3_\parallel H
\right] = 0.
\end{equation}
The characteristic scale for the lateral velocity, $u_\textrm{c}$, is set to the value
established by thermocapillary flow, which can be obtained from
Eq.~(\ref{Eq:surftensgrad_1DTD}) by letting the film thickness, slope and
and interfacial stress be order one and equal to unity - i.e. $H=1$, $\nabla_{\parallel}H =1$,
and $(\partial U/\partial Z)_{Z=H} = \partial \Gamma /\partial X = 1$, such that
\begin{equation}
\label{Eq:surfstress_scaling} \overline{Ma} = \frac{(D_o + \kappa -
1)^2}{\kappa D_o}.
\end{equation}
Since $\overline{Ma} = \epsilon \gamma_T \Delta T/(\eta u_\textrm{c})$, the
scale for $u_\textrm{c}$ becomes
\begin{equation}
\label{Eq:u_c} u_\textrm{c} = \frac{\epsilon\, \kappa \,D_o\, \gamma_T \,\Delta
T}{\eta \,(D_o + \kappa - 1)^2}.
\end{equation}
%

\begin{center}
  \begin{table}
   \citestyle{plain}
      \caption{Order of magnitude estimates for characteristic numbers used in the
      thermocapillary model extracted from the experiments of Sch\"affer \etal
      \cite{SchaefferSteiner:EurophysLett2002,SchaefferSteiner:Macromolecules2003}.
       Values of \textit{Pr} for PS and PMMA at 170$^o$C were obtained from
        Refs. {\citestyle{plain}\cite{Mark:AIPPress1996}} and {\citestyle{plain}\cite{MassonGreen:PhysRevE2002}}.
        The constant value of capillary number, $\overline{Ca}$, results from the
        choice of thermocapillary velocity used to scale the flow speed,
        as discussed in the section following Eq.~(\ref{Eq:tchar_thermocap})}
 \label{Tbl:char_numbers}
    \begin{tabular}{l c | c}
      \hline \hline
        $\epsilon$ & & $10^{-3}\!-\!10^{-2}$ \\
        $Re$ & & $10^{-18} \!-\! 10^{-17}$ \\
        $Pr$ & & $10^8 \!-\! 10^9$ \\
        $\overline{Ma}$ & & $10^0\!-\!10^1$ \\
        $\overline{Ca}$ & & $52.6$ \\
        $\overline{Bo}$ & & $10^{-5}\!-\!10^{-6}$ \\ \hline \hline \\
    \end{tabular}
  \end{table}
\end{center}

\begin{center}
  \begin{table}
    \caption{Literature values for air and polystyrene melt
    [$M_n \approx 107\:\textrm{kg/mol}$ and $M_w/M_n = 1.07$
    where $n$ and $w$ denotes number avg and weight avg] used
    in the analysis and numerical simulations.
    All parameter values quoted are for  $T=170\:^\textrm{o}\textrm{C}$,
    except for $\gamma$ and $\gamma_T$, which were only available
    for $T=180\:^\textrm{o}\textrm{C}$.
    For comparison, Sch\"affer \etal
    \cite{SchaefferSteiner:EurophysLett2002,SchaefferSteiner:Macromolecules2003}
    used polystyrene melts for which $M_n \approx 108\:\textrm{kg/mol}$ and $M_w/M_n = 1.03$. \\
    }
    \label{Tbl:material_properties}
    \citestyle{plain}
    \begin{tabular}{l c l | c l c l}
      \hline \hline
        & & & &  Air & & PS \\ \hline
        $\rho$ & & (kg/m$^3$) & & $0.829\:$\cite{Lide:CRCPubComp1992} & & $987\:$\cite{Mark:AIPPress1996} \\
        $\eta$ & & (Pa$\cdot$ s) & & $2.48\cdot10^{-5}\:$\cite{Lide:CRCPubComp1992} & & $2.5\cdot10^{4}\:$\cite{MassonGreen:PhysRevE2002} \\
        $k$ & & [W/(m$\:^\textrm{o}\textrm{C}$)] & & $0.036\:$\cite{Lide:CRCPubComp1992} & & $0.130\:$\cite{Mark:AIPPress1996} \\
        $\alpha$ & & (m$^2$/s) & & $4.25\cdot10^{-5}\:$\cite{Lide:CRCPubComp1992} & & $6.45\cdot10^{-8}\:$\cite{Mark:AIPPress1996} \\
        $\gamma$ & & ($10^{-3}\:$N/m) & & & & $31.53\:$\cite{MoreiraDermarquette:JAPolymScie2001} \\
        $\gamma_T$ & & [$10^{-3}\:$N/(m$\:^\textrm{o}\textrm{C}$)] & & & & $0.0885$ \cite{MoreiraDermarquette:JAPolymScie2001} \\ \hline \hline \
    \end{tabular}
  \end{table}
\end{center}

The evolution of film disturbances governed by thermocapillary effects,
as given by Eq. (\ref{Eq:TCinterface}), is compared to evolution by acoustic
phonon radiation pressure, as proposed by Sch\"affer \etal. While their derivation is also based on the slender
gap approximation, the acoustic phonon model neglects altogether any flow induced by
tangential stresses due to interfacial thermal gradients.
Instead, the Laplace pressure, is counteracted
by a radiation pressure due to phonon reflections which causes protrusions to grow. The
overall fluid pressure in the AP model is therefore given by
\begin{equation}
\label{Eq:normal_stress_augment} P = -\:\overline{Ca}\:^{-1} \nabla^2_\parallel
H -\overline{Ca}\:^{-1}\overline{Q}/[D_o + (\kappa - 1)H],
\end{equation}
where $\overline{Q} = 2Q k_{\textrm{air}} \Delta T/(u_p \gamma \epsilon^2)$ and
$Q$ is the acoustic quality factor described in the Introduction. Substitution
of Eq. (\ref{Eq:normal_stress_augment}) into Eq. (\ref{Eq:interface}) (with
$\nabla_\parallel \Gamma =0$ since thermocapillary effects play no role in the
acoustic phonon model) yields the evolution equation proposed by Sch\"affer \etal:.
\begin{equation}
\label{Eq:APinterface} \frac{\partial H}{\partial \tau} + \nabla_{\parallel}
\cdot \left[ \frac{\overline{Q}~(1 - \kappa ) H^3}{3~\overline{Ca}~[D_o +
(\kappa - 1 )H]^2} \nabla_{\parallel} H + \frac{H^3}{3~\overline{Ca}}
\nabla^3_{\parallel} H \right]=0.
\end{equation}

Values for the thermophysical properties of air and PS
are listed in Table~\ref{Tbl:material_properties}. Corresponding numbers for
experiments with PMMA \cite{PengHan:Polymer2004} are of similar magnitudes.

\subsubsection{Linear stability analysis of evolution equation}

Equations (\ref{Eq:TCinterface}) and (\ref{Eq:APinterface}) can be further
analyzed by linear stability theory to provide an estimate of the fastest
growing mode, the one most likely to be observed in experiment.
Predictions of the corresponding wavelength are therefore expected to compare
favorably with the pillar spacing measured in experiment if the proposed mechanism
is correct.

The behavior of Eq. (\ref{Eq:TCinterface}) is examined in the limit
where an initially flat and uniform film
of thickness $H=1$ (i.e. base state) is subject to an infinitesimal periodic perturbation
of amplitude $\widetilde{\delta H_o} \ll 1$ and wave number
$\overrightarrow{K}_\parallel$ where $|\overrightarrow{K}_\parallel| = K= 2\pi
L/\lambda$. Solutions of the form $H(X,Y,\tau) = 1 +
\widetilde{\delta H_o} \:\exp[\beta(K)\tau] \exp[i \overrightarrow{K}_\parallel
\cdot \overrightarrow{X}_\parallel]$ are substituted into Eq. (\ref{Eq:TCinterface}), where
$\overrightarrow{X}_\parallel = (X,Y)$, and all quadratic or higher
order terms are neglected. The resulting expression for the growth rate is
\begin{equation}
\label{Eq:growthrate} \beta(K) = \left(\frac{\kappa D_o \overline{Ma}}{2(D_o +
\kappa - 1)^2} -\frac{K^2}{3 \overline{Ca}}\right) K^2.
\end{equation}
Disturbances for which $\beta(K)=0$ neither grow nor decay. This condition
establishes the criterion for marginal (M) stability where the corresponding
wave number, $K_\textrm{M}$, for the thermocapillary model, is given by
\begin{equation}
\label{Eq:onset_Kc} K^{\textrm{TC}}_\textrm{M} = \sqrt{\frac{3}{2} \frac{\kappa
D_o \overline{Ma}\:\overline{Ca}}{(D_o + \kappa -1)^2}}.
\end{equation}
Note that in the absence of any stabilizing hydrostatic terms as is the case
with nanofilms, there always exists a band of wavenumbers $0 < K <
K^{\textrm{TC}}_\textrm{M}$ for which the film is linearly unstable, no matter
how small the value of the imposed temperature gradient. This stands in sharp
contrast to the thermocapillary instability in much thicker films
\cite{VanHook:PRL1995,VanHook:JFM1997} for which $K_\textrm{M} =
(3\:\kappa\:D_o\:\overline{Ma}\:\overline{Ca}/[2(D_o + \kappa -1)^2] -
\overline{Bo})^{1/2}$. For thicker films, there exists a critical Marangoni number for
onset of instability:
\begin{equation}
\label{Eq:critical_Ma} \overline{Ma}_\textrm{onset} = \frac{2}{3}
\frac{\overline{Bo}}{\overline{Ca}} \frac{(D_o + \kappa - 1)^2}{\kappa\:D_o}~.
\end{equation}
This criterion is commonly expressed in terms of the inverse dynamic Bond
number $D^{\textrm{dyn}}_\textrm{onset} = \gamma_T \Delta T_{\textrm{film}}/(\rho g h^2_o)
\geq 2/3(1 + F)^{-1}$, where $\Delta T_{\textrm{film}}$ represents the
temperature drop across the liquid layer [and not the temperature drop across
the bilayer as defined in Eq.~(\ref{Eq:critical_Ma})] and $F = (1 -\kappa)/(D_o
+ \kappa - 1)$ is a constant of order one \cite{VanHook:JFM1997}. The regime
investigated by vanHook \etal for films of the order of several hundred microns
corresponds to values of $D^{\textrm{dyn}}_\textrm{onset}$ in the range $10^{-1} - 1$.
By contrast, representative values for
$D^{\textrm{dyn}}$ in the experiments of Sch\"affer \etal and Peng \etal are of
the order of $10^7$. Nanofilms subject to a transverse thermal gradient are therefore
always linear unstable irrespective of the magnitude of $\Delta T$.

The fastest growing wave number is determined from the extremum of
$\beta(K)$ in Eq.~(\ref{Eq:growthrate}), with the result that
$K^\textrm{TC}_{\textrm{max}} = K^\textrm{TC}_\textrm{M}/\sqrt{2} = 2\pi
L/\lambda_{\textrm{max}}$. In dimensional units, the wavelength of the most
unstable mode is given by
\begin{equation}
\label{Eq:TCwavelength} \lambda^\textrm{TC}_{\textrm{max}} = 2\pi h_o
\sqrt{\frac{4 \gamma h_o}{3\:\kappa\:d_o\:\gamma_T\:\Delta T}}
\left[\frac{d_o}{h_o} + \kappa - 1\right].
\end{equation}
This expression provides an estimate of the average spacing between protrusions
undergoing growth by thermocapillary flow. For the nanofilm experiments described earlier,
$h_o \approx \textrm{O}~(100~\textrm{nm})$, $h_o < d_o\lesssim 8\:h_o$ and $\Delta T
\approx 10-50\:^\textrm{o} \textrm{C}$. This leads to predictions of the pillar spacings
ranging from about 2-20 $\mu$m. (More detailed comparison to
experiments will be discussed in Section III.) According to
Eq. ~(\ref{Eq:TCwavelength}), the characteristic lateral spacing between
nanopillars is determined by the initial film thickness, $h_o$, as well as the
gap ratio $D_o= d_o/h_o$, the ratio of the surface tension to the maximum
change in surface tension, $\gamma/(\gamma_T \Delta T)$, and the ratio of
thermal conductivities $\kappa = k_{\textrm{air}}/k_{\textrm{melt}}$. For cases
in which the geometry and material properties are held fixed, a larger thermal
gradient produces more closely spaced pillars. Reversal of the
thermal gradient such that $T_2 < T_1$ should lead to linearly stable films.

\begin{figure}[htp]
\includegraphics[width=7.5cm]{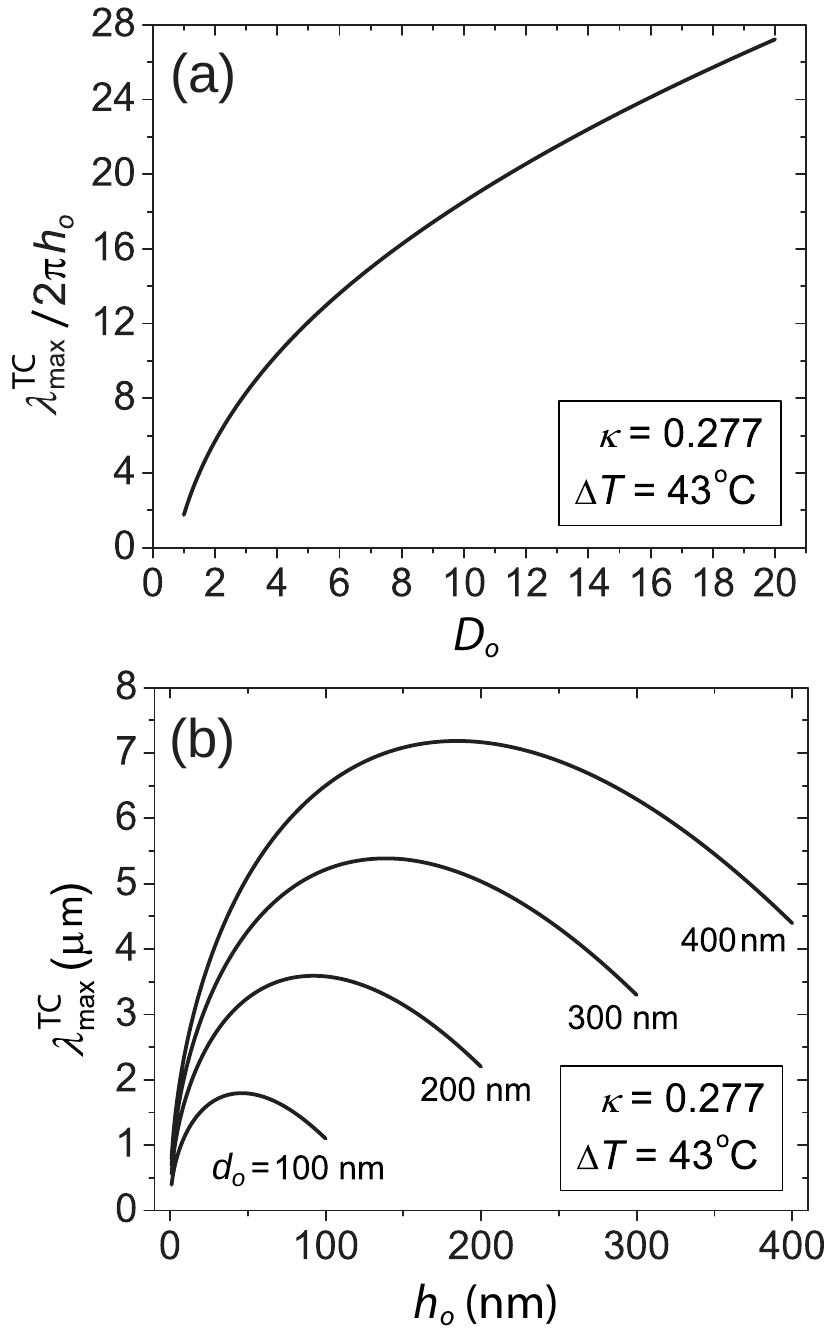}
\caption{Solutions of Eq.~(\ref{Eq:TCwavelength}) for $\kappa = 0.277$ and $\Delta T = 43
^\textrm{o}\textrm{C}$. Curves show a sharp decrease in $\lambda^\textrm{TC}_\textrm{max}$
for the smaller values of $h_o$.}
\label{Fig:lambda_sensitivity}
\end{figure}

Figure \ref{Fig:lambda_sensitivity}(a) represents solutions to Eq.
(\ref{Eq:TCwavelength}) for a polystyrene film at $T_2 = 170 ^\textrm{o}$C
with $\Delta T = 43~^\textrm{o}$C. Smaller gap ratios $D_o$ lead to smaller values of pillar
spacing since the film is subject to a larger effective thermal gradient.
Figure \ref{Fig:lambda_sensitivity}(b) highlights the dependence of $\lambda^\textrm{TC}_{\textrm{max}}$ on the
initial film thickness $h_o$ for various gap widths $d_o$ and $\Delta T=43
^\textrm{o}$C. As evident, the prediction for
$\lambda^\textrm{TC}_{\textrm{max}}$ depends sensitively on $h_o$, especially for
the smallest values of $h_o$.

The linear stability analysis of Eq.~(\ref{Eq:APinterface}) yields a
prediction for the fastest growing wavelength for the acoustic phonon model,
namely Eq. (\ref{Eq:APwavelength}). The ratio of dominant wavelengths
corresponding to the two proposed mechanisms is given by
\begin{equation}
\label{Eq:TCAPratio}
\frac{\lambda^\textrm{TC}_{\textrm{max}}}{\lambda^\textrm{AP}_{\textrm{max}}} =
\sqrt{\frac{4 \,Q \,k_{\textrm{melt}}\,(1-\kappa)}{3\, u_p \,\gamma_T \, D_o}}.
\end{equation}
Future experiments conducted with parallel substrates for a wider range of $D_o$
should help identify the operating mechanism leading to pillar formation.

The characteristic velocity defined earlier in
Eq.~(\ref{Eq:u_c}), which sets the scale for the lateral flow speed based on thermocapillary
stress, can be re-expressed in terms of the length scale
$\lambda^\textrm{TC}_{\textrm{max}}$ obtained from linear stability
analysis:
\begin{equation}
\label{Eq:u_c_tcwavelength}
u_\textrm{c} = \epsilon \frac{(4\pi)^2}{3}
\left(\frac{h_o}{\lambda^\textrm{TC}_{\textrm{max}}}\right)^2
\frac{\gamma}{\eta}= \frac{(4\pi)^2}{3}~\epsilon^3 \left( \frac{\gamma}{\eta} \right).
\end{equation}
Here, the lateral scale $L$ used to define the slender gap parameter,
$\epsilon = h_o/L$, is identified with $\lambda^\textrm{TC}_{\textrm{max}}$.
Similarly, the characteristic timescale based on thermocapillary flow is given
by
\begin{equation}
\label{Eq:tchar_thermocap}
t_\textrm{c}
=\frac{\lambda^\textrm{TC}_{\textrm{max}}}{u_\textrm{c}} = \frac{3 h_o}{(4\pi)^2
\epsilon^4}\left (\frac{\eta}{\gamma} \right).
\end{equation}
Estimates from the experiments of Sch\"affer \etal indicate that $u_\textrm{c}$
is of the order of $10^{-1} - 10^1$ nm/s and
$t_\textrm{c} \approx \textrm{O}(10^{-1} - 10^2\:\textrm{hrs})$. If the
thermocapillary flow speed $u_\textrm{c}$ given by Eq. (\ref{Eq:u_c_tcwavelength}) is used to define
the capillary number, then $\overline{Ca} = (4\pi)^2/3$, a fixed constant. Replacing
the capillary number by this numerical value and substituting the expression
for the Marangoni number given by Eq.~(\ref{Eq:surfstress_scaling}) into the
interface equation Eq.~(\ref{Eq:TCinterface}) yields the following form
of the evolution equation:
\begin{equation}
\label{Eq:TCinterface_simple} \frac{\partial H}{\partial \tau} +
\nabla_\parallel \cdot \left\{\left[\frac{D_o\!+\kappa\!-\!1}{D_o \!+
\!(\kappa\! - \!1)H}\right] ^2\!\frac{H^2}{2} \nabla_\parallel H +\frac{H^3
}{(4\pi)^2} \nabla^3_\parallel H\right\} = 0.
\end{equation}
For thicker films, hydrostatic forces can be re-incorporated into this expression by
including the term $-\overline{Bo}\:H^3 \nabla_\parallel H/(4 \pi)^2$ in the curly brackets.
During the early stages of film deformation when $H$ and $\nabla_\parallel H$ are order one, the
relative magnitude of terms in Eq.~(\ref{Eq:TCinterface_simple})
reveals the basis for pillar formation. The ratio of
thermocapillary to capillary flux scales as $8\pi^2$, while the ratio
of thermocapillary to gravitational flux scales as
$8 \pi^2/\overline{Bo} \approx 10^{7} - 10^{8}$. These estimates reveal that
thermocapillary forces overcome the stabilizing effect of capillary and gravitational forces
even at early times. In section IV.B, it is shown that thermocapillary forces prevail even more
strongly at late times for parameter values pertinent to the nanofilm experiments.
A similar comparison can be made using the parameter values in the experiments of VanHook \textit{et al.}
\cite{VanHook:PhD1996} with thicker films ($70 \lesssim h_o \lesssim 270\:\mu \textrm{m}$)
and much smaller transverse thermal gradients ($180 \lesssim T_2 - T_1/d_o \lesssim 500$ $^\textrm{o}$C/cm).
While the thermocapillary to capillary flux ratio remains at $8\pi^2$,
the thermocapillary to gravitational flux ratio decreases to $10^{-1}$, eight to nine orders of magnitude
smaller than the ratio in the nanofilm experiments of Sch$\ddot{a}$ffer
\textit{et al.} and Peng \etal. While gravitational forces
effectively repress the growth of pillars in macroscopically thick films, this order of magnitude analysis
confirms that hydrostatic forces are ineffective in repressing the growth of elongations in nanoscale films.

Integration of the full nonlinear Eq. (\ref{Eq:TCinterface}) can be used to
compute a lower bound on the time interval, $t_\textrm{top}$, required for nanopillars to contact the
cooler substrate within the approximation of a constant film viscosity \cite{footnote1}.
It will be shown in Section IV.A that estimates obtained from the growth rate of the
most unstable mode, $\beta(K_{\textrm{max}})$, are in fairly good agreement
with the estimates obtained from numerical solutions of
Eq.~(\ref{Eq:TCinterface}) for the parameter range of interest. Substitution of
Eq.~(\ref{Eq:surfstress_scaling}) and $\overline{Ca} = (4\pi)^2/3$ into
Eq.~(\ref{Eq:growthrate}) and Eq.~(\ref{Eq:onset_Kc}) yields the
simplified expression for the growth rate:
\begin{equation}
\label{Eq:growthrate_simplified} \beta^\textrm{TC}(K) = [1/2 -
(K/(4\pi))^2]K^2.
\end{equation}
The wave number corresponding to marginal stability is therefore
$K^\textrm{TC}_\textrm{M} = 4\pi/\sqrt{2}$. Since $K^\textrm{TC}_{\textrm{max}}
= K^\textrm{TC}_\textrm{M}/\sqrt{2} = 2\pi$, the growth rate for the fastest
growing mode simply reduces to $\beta^\textrm{TC}_{\textrm{max}} =\pi^2$.
Setting $\widetilde{\delta H_o} \exp[\beta(K_{\textrm{max}})\tau] = D_o - 1$
leads to the expression $\tau_{\textrm{top}} = \ln[(D_o-1)/\widetilde{\delta H_o}/\pi^2$,
which in dimensional units corresponds to $t_{\textrm{top}} =
(3\:\eta\:h_o/\gamma) [\lambda^{\textrm{TC}}_{\textrm{max}}/(2\pi h_o)]^4
\ln[(D_o-1)/\widetilde{\delta H_o}]$. Substitution of
Eq.~(\ref{Eq:TCwavelength}) into this expression then gives
\begin{equation}
\label{Eq:t_top}
t_{\textrm{top}} = \frac{16\:\eta\:\gamma\:h_o(D_o + \kappa
-1)^4}{3\:(\kappa\:D_o\:\gamma_T\:\Delta T)^2}
\ln\left(\frac{D_o-1}{\widetilde{\delta H_o}}\right).
\end{equation}
Estimates of $t_\textrm{top}$ for the nanopillar experiments range
from about tens of minutes to tens of hours for the largest gap spacings used
and $\widetilde{\delta H_o} = 10^{-5}$.
Low molecular weight polymers with much smaller viscosities
require proportionally less time to contact the cooler top
substrate. Studies of this sort are useful in determining when to remove the
thermal gradient in order to form nanopillars of specified height.

\subsubsection{Lyapunov free energy for evolving interface}

Hydrodynamic systems subject to interfacial instability sometimes
exhibit steady states as observed in Rayleigh-B\'{e}nard or
B\'{e}nard-Marangoni cellular convection.
Within the context of the experiments described, this would require
pillar formations which once formed, neither grow nor decay, representing a
fixed spatial configuration while the melt continues to undergo surface
and interior flow.
To examine this possibility, one can investigate the
temporal behavior of the Lyapunov free energy associated with the evolving interface,
as previously implemented in Refs.
{\citestyle{plain}\cite{Oron:PoF2000,OronRosenau:JFM1994}}. This approach
is based on the analysis of interface problems using the well known form of the
Cahn-Hilliard free energy for systems with spatial variation
in an intensive scalar variable like composition or density \cite{CahnHilliard:JCP1958}.
The Cahn-Hilliard equation has been successfully used to explore
the evolution of moving interfaces in
binary systems undergoing phase separation. This approach, which
involves monitoring the free energy associated with the entire film undergoing
deformation, provides a more accurate assessment of possible steady state configurations
than simple considerations based on Eq.(\ref{Eq:TCinterface_simple})
in the limit $\partial H/\partial \tau \rightarrow 0$.

In the Appendix, it is shown that
the free energy corresponding to the nanofilms of interest is given by
$\mathfrak{F} = \int \mathfrak{L} ~ dX dY$, where
\begin{equation}
\label{Eq:lyapunov} \mathfrak{L}\!=\!(\nabla_{\parallel} H)^2 \!- \!\frac{3
\kappa \overline{Ma}\: \overline{Ca}}{D_o} \left[ H \textrm{ln} \left
(\frac{H}{1 + \chi H} \right) \!+\!\textrm{ln} (1+\chi) \right]
\end{equation}
and $\chi = (\kappa - 1)/D_o$. Numerical solutions of
Eq. (\ref{Eq:lya_final_1}) for large and small values of the gap ratio, $D_o$,
are discussed in Section IV.B.

\subsection{Films confined by non-parallel substrates}

The analysis presented in Section II.A describes the evolution of a fluid
bilayer interface confined by two flat and parallel substrates
separated by a distance $D_o=d_o/h_o$. As described in Section I.A.2, however,
Sch\"affer and co-workers purposely used in all their experiments a
tilted plate geometry in which the top wafer was inclined with respect
to the bottom one by about $1 \mu\textrm{m}/
1 \textrm{cm}$, corresponding to an inclination angle $\varphi$ of about $0.0057 ^\textrm{o}$.
The evolution equation can be modified to account for two flat substrates
with relative tilt. When the cooler substrate is tilted away from
the horizontal by a constant angle $\varphi$, the local value of the plate
separation will depend on $(X,Y)$ such that $D(X,Y) = d(x,y)/h_o$.
This modification alters the film surface temperature, $\Theta |_{Z=H}$, as
well as the surface thermal gradient, $\nabla \Theta |_{Z=H}$, which in
turn alters the interfacial thermocapillary stress,
$\nabla_\parallel \Gamma$. Accordingly,
\begin{equation}
\label{Eq:interfacetemptilt} \Theta |_{Z=H} = (D-H)/[D+(\kappa - 1)H]~,
\end{equation}
\begin{equation}
\label{Eq:interfacetempgradtilt} \nabla_\parallel \Theta |_{Z=H} =
\frac{\kappa}{[D+(\kappa - 1)H]^2} \left( H \nabla_\parallel D - D
\nabla_\parallel H \right),
\end{equation}
and
\begin{eqnarray}
\label{Eq:surftensgrad_1DTD_tilt}
\nabla_\parallel \Gamma |_{Z=H} &=& - \overline{Ma} \nabla_\parallel \Theta |_{Z=H} \nonumber \\
&=& \kappa \overline{Ma} ~ \frac{\left(D \nabla_\parallel H -
H\tan(\overline{\varphi})\overrightarrow{T}_\parallel
\right)}{[D+(\kappa-1)H]^2}.
\end{eqnarray}
Here, $D(\overrightarrow{X}_\parallel) = D_o +
\tan(\overline{\varphi})\overrightarrow{T}_\parallel \cdot
\overrightarrow{X}_\parallel$, where $D_o$
represents the gap ratio at $\overrightarrow{X}_\parallel = 0$ (later
identified with the midpoint of the computational domain). The quantities
$\overline{\varphi}$ and $\tan \overline{\varphi} = \tan \varphi/\epsilon$
represent variables rescaled according to the slender gap approximation. In the
numerical solutions discussed in Section IV.C.2, the tilt of
the upper substrate is defined by the unit vector $\overrightarrow{T}_\parallel
= (1,1)/\sqrt{2}$. Substitution of Eq.~(\ref{Eq:surftensgrad_1DTD_tilt}) into
Eq.~(\ref{Eq:nstressgrad}) and Eq.~(\ref{Eq:interface}) leads to the modified
evolution equation
\begin{equation}
\label{Eq:TCinterface_tilt} \frac{\partial H}{\partial \tau} + \nabla_\parallel
\cdot \vec{Q}_\textrm{tilt} = 0~,
\end{equation}
where
\begin{equation}
\vec{Q}_\textrm{tilt} = \frac{\kappa \overline{Ma} H^2(D \nabla_\parallel H -
H\tan(\overline{\varphi})\overrightarrow{T}_\parallel)}{2[D + (\kappa - 1)H]^2}
+\frac{H^3}{3 \overline{Ca}} \nabla^3_\parallel H.
\end{equation}
A linear stability analysis of
Eq.~(\ref{Eq:TCinterface_tilt}) (not shown here) confirms that the pattern wavelength
in Eq.~(\ref{Eq:TCwavelength}) remains unaffected by the
small tilt angle used in the experiments of Sch\"affer \etal. More
generally, Eq.~(\ref{Eq:TCwavelength}) remains valid so
long as $|\tan(\overline{\varphi})| \leq \textrm{O}(\widetilde{\delta H_o})$.

\section{Nanopillar Spacings: Comparison Between Experiment and Theory}

\begin{figure}[htp]
\includegraphics[width=7.5cm]{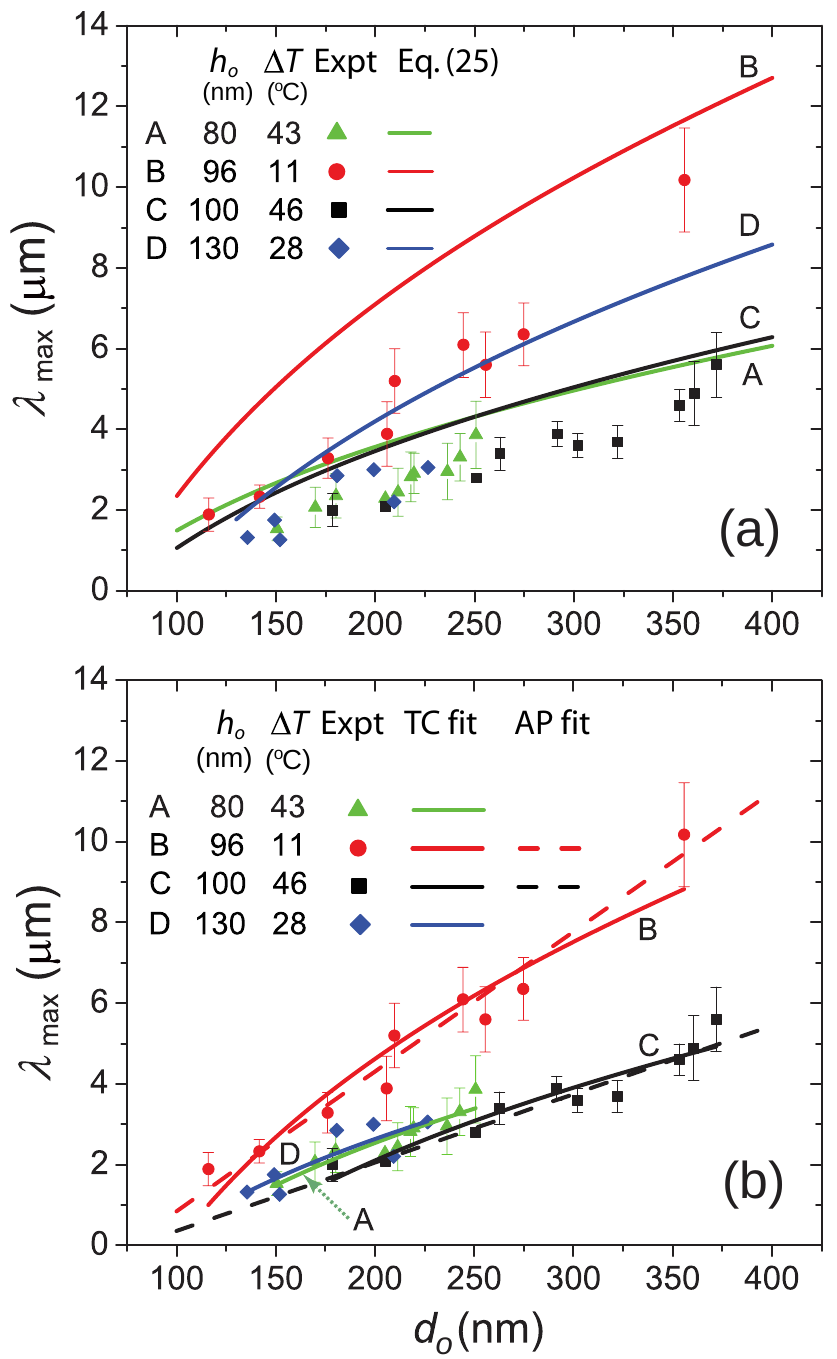}
\caption{Direct comparison of theoretical estimates for dominant
instability wavelength, $\lambda_{\textrm{max}}$ for both thermocapillary
(TC) and acoustic phonon (AP) models with experimental measurements from
Sch\"affer \etal
\cite{SchaefferSteiner:EurophysLett2002,SchaefferSteiner:Macromolecules2003,
SchaefferSteiner:AdvMat2003,Schaeffer:UKonstanz2001} as function of
increasing wafer separation distance $d_o$. (a) Plots of
Eq.~(\ref{Eq:TCwavelength}) for thermocapillary model with no adjustable
parameters for different experiments labeled A-D using material
constants listed in Table~\ref{Tbl:material_properties}. (b) Plots showing
least squares fits to the TC and AP models. The TC model was fitted to the form
given by Eq.~(\ref{Eq:TCwavelength}), namely
$\lambda^{\textrm{TC}}_{\textrm{max}} = C_1 \sqrt{d_o} + C_2/\sqrt{d_o}$.
AP model was fitted to the function given by Eq.~(\ref{Eq:APwavelength}) with
$u_p = 1850\:\textrm{m/s}$ and $Q = 6.2$ as fitting parameters.
Table~\ref{Tbl:datafit_thermocap_REV} lists the fitting coefficients, $C_1$ and
$C_2$, obtained for the TC model.}
\label{Fig:comp_exptheo}
\end{figure}

Shown in Fig.~\ref{Fig:comp_exptheo}(a) is a direct comparison of
Eq.~(\ref{Eq:TCwavelength}) with the experimental data of Sch\"affer \etal
\cite{SchaefferSteiner:EurophysLett2002,SchaefferSteiner:Macromolecules2003}
The solid lines denote the predictions of the thermocapillary model with no
adjustable parameter values using the material properties listed in Table
\ref{Tbl:material_properties}; the symbols denote the experimental data.

While the overall functional behavior of $\lambda^\textrm{TC}_{\textrm{max}}$ with
$d_o$ is in good agreement with experiment, the model
systematically overestimates the pillar spacings, in some
cases by as much as $40\%$. This is especially  evident in experimental
run B for which $h_o = 96$ nm and $\Delta T = 11\:^\textrm{o} \textrm{C}$.
Before discussing these discrepancies in detail, it is useful to
examine a least-squares fit of the data to the function
$\lambda^\textrm{TC}_{\textrm{max}}=C_1\:{d_o}^{1/2} + C_2\:{d_o}^{-1/2}$
given by Eq.~(\ref{Eq:TCwavelength}), as shown in
Fig.~\ref{Fig:comp_exptheo}(b). Listed in Table~\ref{Tbl:datafit_thermocap_REV}
is a comparison of the analytic expressions for the two constants, namely
$C^\textrm{TC}_1 = 2\pi[{4\:h_o\:\gamma/(3\:\kappa\:\gamma_T\:\Delta T]^{1/2}}$
and $C^\textrm{TC}_2 = C_1\:(\kappa - 1)\:h_o$, along with the results for the
fitting constants denoted by $C^{\textrm{Fit}}$.

\begin{center}
  \begin{table}
  \citestyle{plain}
    \caption{Coefficients obtained from a least squares fit to the data of Sch\"affer \etal
    \cite{SchaefferSteiner:EurophysLett2002,SchaefferSteiner:Macromolecules2003}.
    Experimental data were fit to the function $\lambda^{\textrm{TC}}_{\textrm{max}}= C_1\:{d_o}^{1/2} + C_2\: {d_o}^{-1/2}$
    given by Eq.~(\ref{Eq:TCwavelength}), where
    $\lambda^{\textrm{TC}}_{\textrm{max}}$ is reported in microns and
    $h_o$ and $d_o$ in nm. The constants
    $C^\textrm{Fit}$ denote the values obtained for the least squares fits shown in Fig. \ref{Fig:comp_exptheo}.
    The constants $C^\textrm{TC}$ represent the predictions of the TC model given
    by Eq.~(\ref{Eq:TCwavelength}) with no adjustable parameters using the material properties listed in
    Table~\ref{Tbl:material_properties}. The percentage errors are defined by
    $(C^\textrm{Fit} -~C^\textrm{TC})/C^\textrm{TC}$.\\
    }
    \label{Tbl:datafit_thermocap_REV}
     \begin{tabular}{c c | c c c c}
      \hline \hline
      & & A & B & C & D \\
      & $h_o ~~~(\textrm{nm})$  & 80 & 96 & 100 & 130 \\
      & $\Delta T ~~~(^\textrm{o}\textrm{C})$ & 43 & 11 & 46 & 28 \\ \hline
      $C^\textrm{TC}_1$ & [$(10^{3}\:\mu \textrm{m})^{0.5}$] & 0.36 & 0.77 & 0.38 & 0.56 \\
      $C^\textrm{TC}_2$ & [$(10^{-1}\:\mu \textrm{m})^{1.5}$] & -21 & -53 & -28 & -53 \\ \hline
      $C^\textrm{Fit}_1$ & [$(10^{3}\:\mu \textrm{m})^{0.5}$] & 0.35 & 0.65 & 0.38 & 0.34 \\
      && $\pm$ &$\pm$ &$\pm$ &$\pm$\\
      & & 0.036 & 0.058 & 0.031 & 0.071 \\
      \hline
      $C^\textrm{Fit}_2$ & [$(10^{-1}\:\mu \textrm{m})^{1.5}$] & -35 & -65 & -46 & -31 \\
      && $\pm$ &$\pm$ &$\pm$ &$\pm$\\
      & & 7.5 & 12 & 8.7 & 13 \\
      \hline
      $\%$ Error $C_1$ &  & -0.53 & 16 & 1.2 & 39 \\
      $\%$ Error $C_2$ &  & -69 & -21 & -66 & 42 \\ \hline \hline \\
    \end{tabular}
  \end{table}
  \end{center}

In general, the agreement between the TC model and experiment improves for larger values of $\Delta T$.
However, given that the least squares fit
captures the experimental trend with increasing values of $h_o$ and $d_o$ so well,
it is worth considering what experimental challenges might affect the reported measurements.
For completeness, we include in Fig.~\ref{Fig:comp_exptheo}(b) two additional dashed lines
for runs B and C,
which represent a least squares fit of the data to
Eq.~(\ref{Eq:APwavelength}) with $Q=6.2$ and $u_p$ = 1850 m/s, the same fitting
constants reported by Sch\"affer \etal
\cite{SchaefferSteiner:EurophysLett2002,SchaefferSteiner:Macromolecules2003}

\subsection{Possible causes of discrepancy between theory and experiment}
There are several experimental challenges in performing the experiments on
nanopillar formation. Perhaps the most important is that all experiments to date
have used silicon wafers to confine the polymer films. These opaque substrates
prevent observation of the instability \textit{in-situ}. In fact, measurements
of the pillar spacings were normally obtained long after the pillars
had contacted the cooler wafer. The pillar amplitudes were by then sizeable,
possibly violating the assumptions of linear stability analysis. Furthermore,
the warmer nanopillars had sustained prolonged contact
with a cooler substrate leading to possible reorganization of fluid
due to thermocapillary or other packing effects along the underside of the
top wafer. Measurements taken
once the pillars had solidified and the top wafer was removed may therefore differ
from the predictions of linear stability theory. In many of the experiments described earlier,
measurements of the spacing between fluid elongations included not only pillar arrays, but
lamellar, spiral and other periodic structures since these were more commonly obtained.
An additional complication is that a typical molten nanofilm is not completely smooth and flat
due to the presence of contaminant particles and pinholes caused by dewetting. Any small fluid elevations
caused by these nucleation points are prone to rapid growth when subject to a thermal gradient.
Structures arising from such initial conditions, however, correspond more to disturbances of finite amplitude and
not infinitesimal amplitudes as assumed by the linear analysis.

As evident from the curves in Fig.
~\ref{Fig:lambda_sensitivity}, the parameters $h_o$ and $d_o$ strongly affect the
predicted values of $\lambda^\textrm{TC}_{\textrm{max}}$. The
sharp drop in $\lambda^\textrm{TC}_{\textrm{max}}$ becomes even more
pronounced for smaller values of $\Delta T$ \cite{DietzelTroian:PRL2009}. Validation
of either mechanism proposed therefore requires accurate measurements of the film thickness.
It appears that the films used by Sch\"affer \etal
\cite{Schaeffer:UKonstanz2001,SchaefferSteiner:EurophysLett2002,SchaefferSteiner:Macromolecules2003,SchaefferSteiner:AdvMat2003}
and Peng \etal \cite{PengHan:Polymer2004} were not annealed prior to insertion
in the experimental setup.
Spun cast polymer films tend to retain a significant amount of
solvent \cite{GarciaJerome:CollPolym2007,PerlichMuller:Macromol2009}, which is
normally expelled by film annealing in vacuum at elevated temperatures
for several hours. (Annealing has the additional advantage of healing pin
holes that sometimes form during spin coating.) Significant film shrinkage typically
accompanies this process due to solvent evaporation. The degree of film shrinkage depends on
the ambient vapor pressure as well as the time and temperature of the
bake. It is therefore likely that the values of
$h_o$ reported in the literature represent overestimates of the initial film thickness $h_o$.
Smaller values of $h_o$ lead to smaller predictions for the pillar
spacing, in closer agreement with experiment.

The distance between pillars in experiment was typically obtained by direct
measurement from optical micrographs. In future experiments, it would be preferable
to Fourier analyze the patterns obtained by an FFT (Fast Fourier Transform) analysis.
This analysis may reveal not only the dominant wave number but
harmonics that develop due to the growth of smaller pillars in between two larger neighboring ones.
Such an analysis, however, requires a fair number of protrusions
for statistically meaningful results. It may have been the case with the tilted plate
geometry, that the smaller domains corresponding to each distinct value of
$D_o$ forbade  use of this technique.

We conducted FFT analyses of nanopillar arrays
published in the literature
\cite{Schaeffer:UKonstanz2001,SchaefferSteiner:EurophysLett2002,SchaefferSteiner:Macromolecules2003}
and were surprised to find a very wide distribution in pillar spacings within even a
single experiment. Often there appeared not a single dominant
wavelength but several competing wavelengths. This finding prompted a
sensitivity analysis of Eq.~(\ref{Eq:TCwavelength}) to better understand which
variables most strongly affect the uncertainty in
measurements of $\lambda^\textrm{TC}_{\textrm{max}} =
\lambda^\textrm{TC}_{\textrm{max}}(\xi_i)$, as defined by
$U_{\lambda^\textrm{TC}_{\textsf{max}}} =
\sqrt{\sum_{i}(S_{\xi_i}\:\Delta\xi_i/\xi_i)^2}$. Here, the relative
sensitivity coefficients are given by $S_{\xi_i}\!=\!\xi_i\:\partial
\lambda^\textrm{TC}_{\max}/\partial \xi_i$ where $\xi_i =
(\gamma,\:\gamma_T,\:\Delta T,\:\kappa,\:D_o,\:h_o)$. This analysis
demonstrates that $S_{\gamma} = -S_{\gamma_T} = -S_{\Delta T} =
1/2\:\lambda^\textrm{TC}_{\textrm{max}}$, $S_{\kappa} = [\kappa/(D_o + \kappa -
1) - 1/2]\lambda^\textrm{TC}_{\textrm{max}}$, $S_{D_o} = [D_o/(D_o + \kappa -
1) - 1/2]\lambda^\textrm{TC}_{\textrm{max}}$ and $S_{h_o} = [1/2 + (\kappa -
1)/(D_o + \kappa - 1)]\lambda^\textrm{TC}_{\textrm{max}}$. Typical values for
these sensitivity coefficients for the parameter values corresponding to the
experiments of Sch\"affer \etal are summarized in Table~\ref{Tbl:sens_coeff}.
These values indicate that the gap ratio, $D_o = d_o/h_o$, the initial film
thickness, $h_o$, and the polymer surface tension, $\gamma$, most significantly
influence the degree of uncertainty in measurements of
$\lambda^{\textrm{TC}}_{\textrm{max}}$.

\begin{center}
  \begin{table}
      \caption{Typical values of sensitivity coefficients $S_{\xi_i}/\lambda^{\textrm{TC}}_{\textrm{max}}$
      resulting from Eq.~(\ref{Eq:TCwavelength})\\
       where $\xi_i = \gamma,\:\gamma_T,\:\Delta T,\:\kappa,\:D_o ~ \textrm{or}~ h_o$.\\
    }
    \label{Tbl:sens_coeff}
    \begin{tabular}{c|c}
      \hline \hline
        $S_{\gamma}/\lambda^{\textrm{TC}}_{\textrm{max}}$ & 0.5 \\
        $S_{\gamma_T}/\lambda^{\textrm{TC}}_{\textrm{max}},\:S_{\Delta T}/\lambda^{\textrm{TC}}_{\textrm{max}}$ & -0.5 \\
        $S_{\kappa}/\lambda^{\textrm{TC}}_{\textrm{max}}$ & -0.5 to -0.25 \\
        $S_{D_o}/\lambda^{\textrm{TC}}_{\textrm{max}}$ & 0.6 to 1.1 \\
        $S_{h_o}/\lambda^{\textrm{TC}}_{\textrm{max}}$ & 0 to 0.4 \\ \hline \hline
    \end{tabular}
  \end{table}
\end{center}

\section{Numerical simulation of thin film equation: linear and non-linear regimes}

To investigate the extent of non-linear effects on the growth of nanopillars,
we also conducted 3D finite element simulations of Eq.~(\ref{Eq:TCinterface}) using
a commercial software package \cite{COMSOL_V3.5a}.
Material properties corresponding to molten polystyrene (PS) were used in these
numerical studies (see Table~\ref{Tbl:material_properties}).
The computational domain corresponded to a square of size
$\Delta X \times \Delta Y = 6 \lambda^\textrm{TC}_{\textrm{max}} \times 6
\lambda^\textrm{TC}_{\textrm{max}}$ (according to Eq.~(\ref{Eq:TCwavelength}))
where spatial discretization was obtained via second order Lagrangian shape functions.
This choice in domain size and discretization order reflects a compromise
between available computational resources and
generation of a sufficient number of peaks for FFT analysis.
Periodic boundary conditions were enforced
along the domain edges (except for the simulations using tilted substrates).
A quadrilateral mesh consisting of $200 \times 200$
elements was applied for the coarse (non-extended) discretization, leading
to an extended  system of equations with about $5\cdot10^5$ degrees of
freedom. An implicit Newton iteration scheme was used to advance the position of the film
interface in time; the linear system of equations for each iteration was solved
using the iterative solver GMRES (Generalized Minimal Residual Method). All
simulations were conducted on HP ProLiant DL360 G4p workstations equipped with
dual Intel Xeon 3.0 GHz processors running CentOS 4.6. The typical growth of a nanopillar
spanning two substrates (i.e. $\tau = \tau_\textrm{top}$) required approximately
$5-6\:\textrm{hrs}$ of CPU time, corresponding to about 900-1000 integration
steps. Numerical convergence tests were conducted by evaluating the local dimensionless film
height at $N=400$ interpolation points within the square domain.
These tests confirmed that both the
average difference,
$\Delta H_\textrm{avg} = \sum^N_{i=1}|H_2(X,Y,\tau_{\textrm{top})} - H_1(X,Y,\tau_{\textrm{top})}|/N$,
as well as the maximum difference,
$\Delta H_\textrm{max} = \max|H_2(X,Y,\tau_\textrm{top}) - H_1(X,Y,\tau_\textrm{top})|/N$, in film
height at the end of a run (i.e. $\tau=\tau_\textrm{top}$)
were less than $10^{-4}$ when decreasing the grid size or
integration time step.
Here, $H_1$ denotes the coarser measurement and
$H_2$ the refined one. Further tests revealed that the film volume was
conserved during each run to a value $\Delta V/V = |\int [H(X,Y,
\tau_{\textrm{top}})- H(X,Y,\tau=0)]~dXdY|/(\Delta X \Delta Y) \leq 10^{-10}$.

In all simulations conducted, the thickness of the initial flat film
was modulated by a very small amount of
white noise such that $H(X,Y,\tau=0) = 1 + \xi\:\mathfrak{R}$, where
$\mathfrak{R}$ denotes a random number between
$-1$ and $+1$. The amplitude of the white noise
was set to $\xi = \textrm{O}(10^{-5})$. According to Eq. (\ref{Eq:t_top}),
larger values of $\xi$ will lead to shorter contact times in proportion
to $-\ln ({\widetilde{\delta H_o}})$. In order to facilitate
detailed comparison between runs for different choices of experimental
parameters, the random number algorithm was reset before each run so as to
generate an identical white noise distribution. Initialization
with white noise was preferable to initialization by a sinusoidal function,
as is common, in order not to bias the system toward a preferred
wavelength too early in the pillar formation process.

\subsection{Films confined by parallel wafers}

\begin{figure}[htp]
\includegraphics[width=7.5cm]{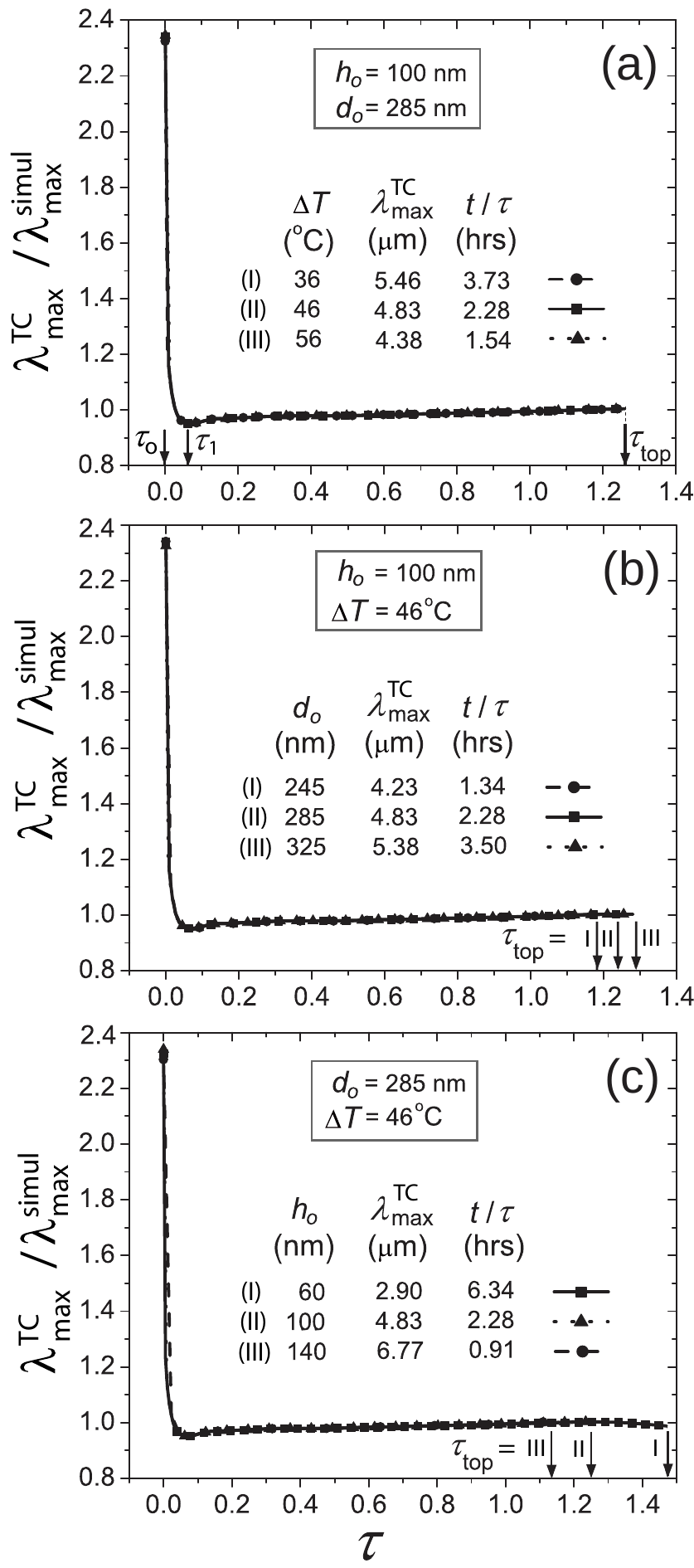}
\caption{Direct comparison of $\lambda^{\textrm{TC}}_{\textrm{max}}$
from Eq.~(\ref{Eq:TCwavelength}) with the instantaneous wavelength,
$\lambda^\textrm{simul}_\textrm{max}$, extracted from FFT analysis of
numerical solutions of the evolving film thickness from
Eq.~(\ref{Eq:TCinterface}) with increasing dimensionless time $\tau$.
Symbol $\tau_\textrm{top}$ represents time of contact of the fastest growing
pillar with cooler substrate. Times $\tau_o=0$, $\tau_1=0.06$ and $\tau_\textrm{top}=1.23$
shown in the top panel refer to time stamps of the snapshot images shown next in
Fig.~\ref{Fig:tfheight_3times}. (a) Variation of the wavelength ratio with
increasing temperature difference $\Delta T$. (b) Variation of the wavelength
ratio with increasing gap separation distance $d_o$. (c) Variation of the
wavelength ratio with increasing values of initial film thickness $h_o$.}
\label{Fig:fft_timeevo}
\end{figure}

FFTs of the in-plane images obtained from the numerical solution of Eq.~(\ref{Eq:TCinterface})
were used to extract values of the dominant wavelength, $\lambda^\textrm{simul}_\textrm{max}(\tau)$,
at each instant in time. This numerical value was
compared to the theoretical prediction
$\lambda^{\textrm{TC}}_{\textrm{max}}$ given by
Eq.~(\ref{Eq:TCwavelength}). Shown in Fig.~\ref{Fig:fft_timeevo} are
results of these simulations. 

\begin{figure}[htp]
\includegraphics[width=8.5cm]{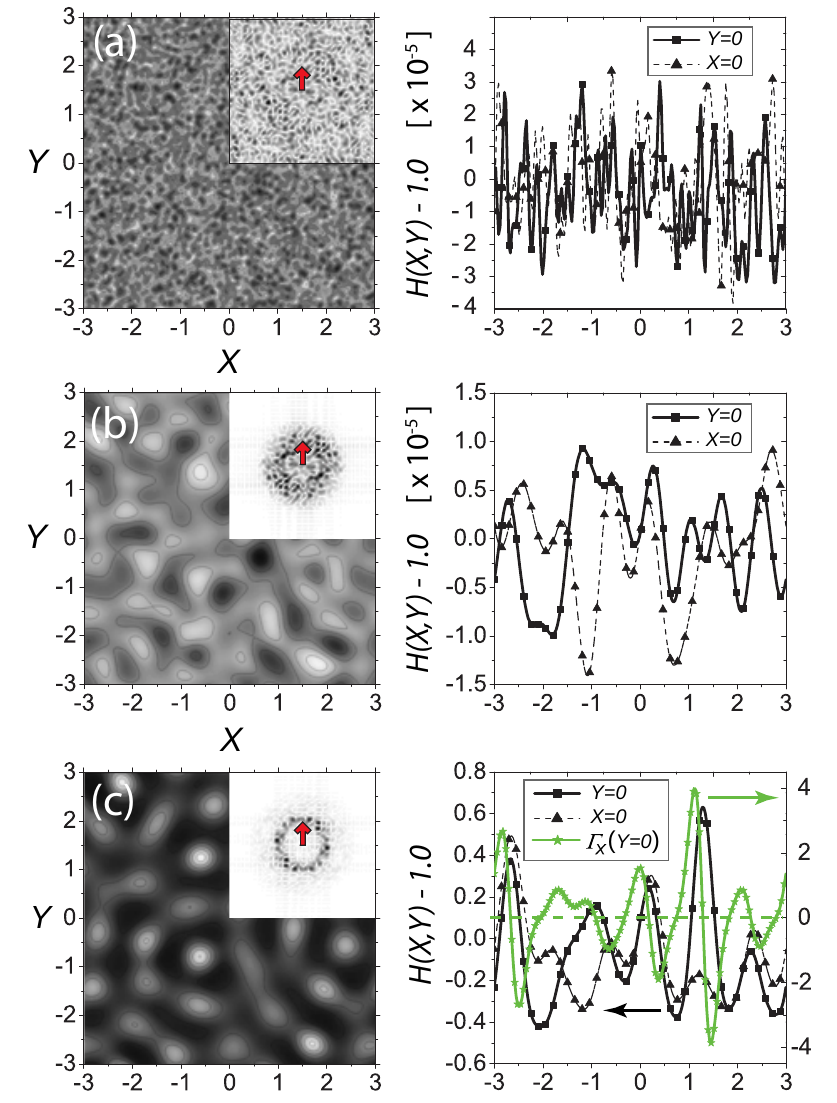}
\caption{View from above of numerical solutions of the film thickness, $H(X,Y,\tau)
-1.0$ from Eq.~(\ref{Eq:TCinterface}) at three instants in time: (a)
$\tau_0 = 0$ (origin of time), (b) $\tau_1=0.06$ and $\tau_\textrm{top}=1.23$.
Simulation parameters values are $h_o$= 100 nm, $d_o$=285 nm and $\Delta T = 46 ^\textrm{o}\textrm{C}$.
Evolution of the corresponding dominant wavelength is depicted by case (\textsf{II})
in Fig.~\ref{Fig:fft_timeevo}(a). Left
panel depicts amplitude $H(X,Y,\tau)-1.0$ (white = elevations, black =
depressions); right panel depicts cross sectional views along axes $X=0$
and $Y=0$. Inset images show the instantaneous 2D Fourier transform of the
corresponding film thickness. Unit arrows denote magnitude of
most unstable wave number, $K^\textrm{TC}_\textrm{max}/\sqrt{2} = 2 \pi$,
derived from linear stability theory [see discussion following Eq.
(\ref{Eq:growthrate_simplified})]. Values of the dimensionless interfacial shear
stress, $\Gamma_X=\partial \Gamma/ \partial X$, along the axis $Y=0$ are shown in the bottom
right image.}
\label{Fig:tfheight_3times}
\end{figure}

The FFTs were computed by sampling $200 \times 200$
points within the computational domain for each value of $\tau$;
approximately 140 instances in time were so evaluated. The legend in each
plot represents the variables held fixed during the simulation; the table entries
specify the theoretical values of
$\lambda^\textrm{TC}_\textrm{max}$ corresponding to the chosen parameter
set. For convenience, the factor used in converting $\tau$ to real time $t$ is
also listed. The times $\tau_o=0$, $\tau_1=0.06$ and $\tau_\textrm{top}=1.23$
shown in Fig.~\ref{Fig:fft_timeevo}(a) denote the three instances in time for
for which the FFTs shown in Fig. ~\ref{Fig:tfheight_3times} were computed. The variable
$\tau_\textrm{top}$ denotes the time at which the fastest growing nanopillar in a
particular run made contact with the cooler substrate, at which point the simulation is terminated.
The times $\tau_\textrm{top}= \textsf{I, II or III}$ indicate this contact time for the
parameters values designated by (\textsf{I}), (\textsf{II}) or (\textsf{III}).

As evident, the overall deviation of $\lambda^\textrm{simul}_\textrm{max}(\tau)$ from
$\lambda^{\textrm{TC}}_{\textrm{max}}$ is rather small regardless of the parameter
range used. In all cases, this ratio rapidly approaches unity
as $\tau \rightarrow 1$. The only discernible difference is that the
fastest growing peaks require a longer time to contact the cooler substrate for
larger values of the relative gap spacing $D_o = d_o/h_o$, as expected.
The very short lived but large initial transients are caused by initialization with
white noise; shortly following $\tau = 0$,
there exist disturbances of all wavelengths. Those
contributions with wave number larger than the cut-off wave number
$K_\textrm{c}$ become rapidly damped. The ratio
$\lambda^\textrm{simul}_\textrm{max}(\tau)/\lambda^{\textrm{TC}}_{\textrm{max}}$
then drops sharply to a value close to one as the maximally unstable disturbance
is established. The approach to unity from below rather than above is due to the
asymmetry in the dispersion curve $\beta(K)$ for which there exists a broader band of
unstable wave numbers below $K_{\textrm{max}}$ than above.

Additional simulations (not shown for brevity) reveal that
$\lambda^\textrm{simul}_\textrm{max}/ \lambda^{\textrm{TC}}_{\textrm{max}}
\rightarrow 1$ by $\tau = 1$ irrespective of the specific initialization
function used i.e.  white noise or a simple sinusoidal function.
Initialization by a double cosine wave in $(X,Y)$ with wavelength
$\lambda_\textrm{c} = \lambda^{\textrm{TC}}_{\textrm{max}}/\sqrt{2}$, for example,
produced the same long time behavior shown so long as the amplitude of the disturbance
function satisfied $\widehat{\delta H_o} \ll 1$.

Images of the evolving film thickness, $H(X,Y,\tau)-1$, as seen from above,
the corresponding Fourier transform (insets), and cross-sectional views
along the mirror planes $X=0$ and $Y=0$ are shown in
Fig.~\ref{Fig:tfheight_3times} at times $\tau = 0,\:0.06 ~ \textrm{and}\:1.23$.
The relevant parameters values are $h_o = 100\:\textrm{nm}$, $d_o = 285\:\textrm{nm}$ and
$\Delta T = 46\:^\textrm{o}\!\textrm{C}$, which represent case \textsf{II}) in Fig.~\ref{Fig:fft_timeevo}].
The arrow shown in the FFT with unit length denotes the magnitude $K^{\textrm{TC}}_{\textrm{max}}$.
As evident from the images
in Figs.~\ref{Fig:tfheight_3times}(a) and (b), although the disturbance heights
of order $10^{-5}$ do not increase substantially from $\tau=0$ to 0.06, an
increasingly regular hexagonal pattern is already visible, both in the
Fourier transform as well the cross sectional views.
Figure~\ref{Fig:tfheight_3times}(c) depicts the in-plane symmetry in a fully
evolved film, just as the fastest growing peak contacts the cooler
substrate. Here, the pillar amplitudes have increased substantially in
comparison to their initial values. By this time, the Fourier transform of the
emerging pattern has evolved from a wide band into a narrow ring with distinct
six-fold symmetry and mean radius $K^{\textrm{TC}}_{\textrm{max}}$.
Values of the (dimensionless) interfacial shear stress, $\Gamma_X =
\partial \Gamma/\partial X$, along the axis $Y=0$ for $\tau = 1.23$ are shown in the bottom right image.
As expected from symmetry, the local extrema in film thickness
along $Y=0$ (solid black curve) occur at the locations of vanishing
shear stress i.e. $\Gamma_X = 0$. The largest values of $|\Gamma_X|$ tend to occur
near the maxima and minima in film thickness.

\begin{figure}[htp]
\includegraphics[width=8.5cm]{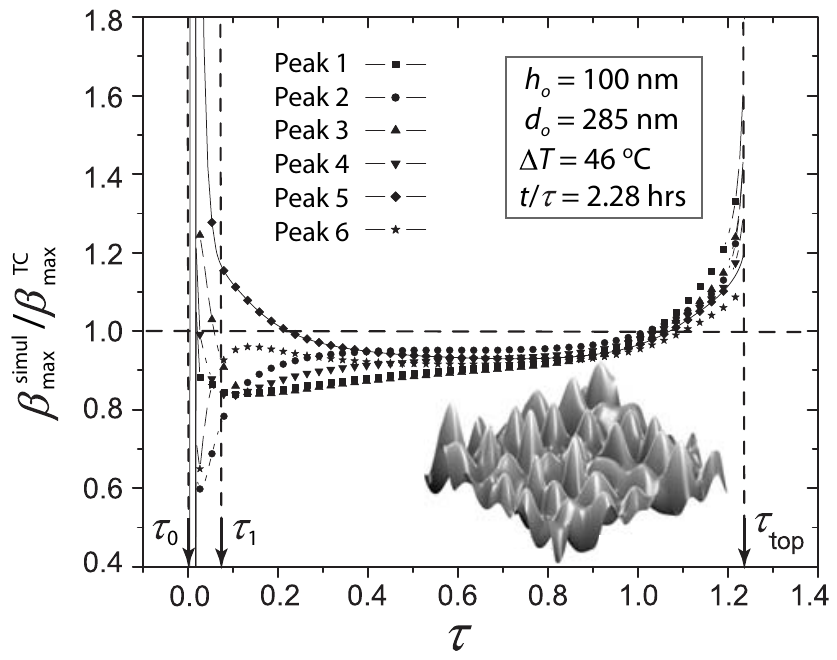}
\caption{Direct comparison of instability growth rate,
$\beta^{\textrm{TC}}_{\textrm{max}}=\pi^2$ (as discussed in Section II.A.1),
with instantaneous growth rate, $\beta^\textrm{simul}_\textrm{max}$, from
Eq.~(\ref{Eq:beta_ratio}), with increasing time $\tau$. Different curves
shown correspond to growth rates of six fastest peaks for
parameter values $h_o\!=\!100$ nm, $d_o\!=\!285$ nm and
$\Delta T\!=\!46\:^\textrm{o}\textrm{C}$. Inset image depicts film shape for
$H(X,Y,\tau_\textrm{top} = 1.23)$.}
\label{Fig:growthrate}
\end{figure}

Shown in Fig.~\ref{Fig:growthrate} is the growth rate ratio,
$\beta^\textrm{simul}_\textrm{max}/\beta^\textrm{TC}_{\textrm{max}}$, for the
parameter values labeled (b) in Fig.~\ref{Fig:fft_timeevo} and
Fig.~\ref{Fig:tfheight_3times}. This ratio was computed for each of the six
most rapidly growing peaks according to
\begin{equation}
\label{Eq:beta_ratio}
\frac{\beta^\textrm{simul}_\textrm{max}}{\beta^\textrm{TC}_{\textrm{max}}} =
\frac{3\:\eta\:h_o}{\gamma}\left(\frac{\lambda^{\textrm{TC}}_{\textrm{max}}}{2\pi
_o}\right)^4 \frac{1}{\delta h_o} \frac{\partial (\delta h_o)}{\partial t}.
\end{equation}
Here, $\beta^\textrm{TC}_{\textrm{max}}=\pi^2$, as shown in Section II.A.1.

As in the solutions shown in Fig.~\ref{Fig:fft_timeevo}, here
too the numerical results are initially influenced by the white noise
disturbance spectrum. Each of the six fastest growing peaks behaves somewhat
differently at the earliest times depending on what is the local value of the
disturbance height. However, the growth rates collapse rapidly
by about $\tau = 0.4$, after which the average growth rate slowly increases
toward the prediction of linear stability theory, which is established by
about $\tau = 1.0$. Beyond this time, the solutions reveal rapid growth and an
increasing departure from the predictions of linear stability theory as
nonlinear effects contribute to the evolving pattern. Beyond $\tau
\approx 1.0$, the growing nanopillars are within reach of the cooler
substrate. The instances marked $\tau_0$, $\tau_1$ and $\tau_{\textrm{top}}$ represent exactly
those times indicated in Fig.~\ref{Fig:fft_timeevo}(a) and
Fig.~\ref{Fig:tfheight_3times}.

\subsection{Numerical simulations of Lyapunov free energy}

\begin{figure}[htp]
\includegraphics[width=7.5cm]{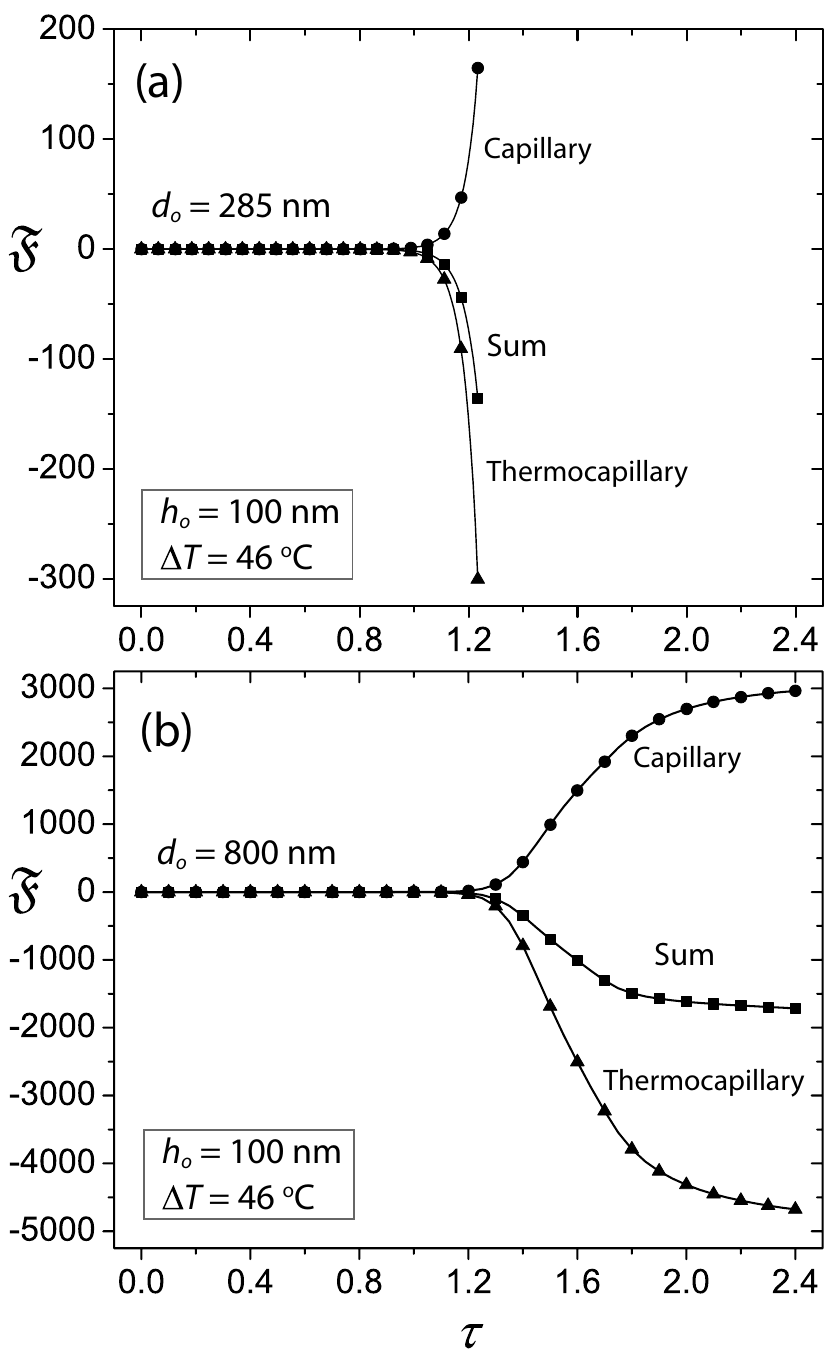}
\caption{Numerical solutions of Lyapunov free energy, $\mathfrak{F}(\tau)$,
for an initial flat film of thickness $h_o = 100$ nm and temperature
difference $\Delta T = 46 ^\textrm{o}\textrm{C}$ subject to two wafer separation distances:
(a) $d_o$ = 285 nm and (b) 800 nm.}
\label{Fig:LyapunovPlot}
\end{figure}

Numerical solutions of Eq.~(\ref{Eq:TCinterface}) confirm that
nonlinear effects for the parameter sets examined
become signficant only when fluid elongations come into close
proximity with the cooler substrate. As evident in Fig. \ref{Fig:growthrate},
the elongation rate then exceeds exponential growth. In this regime, the
nanopillars have grown a distance large in comparison to the initial film
disturbance heights and the nonlinear terms in
Eq.~(\ref{Eq:TCinterface}) strongly influence the flow.
To explore the energetics of formation beyond the linear regime, we investigated the temporal
behavior of the Lyapunov free energy given by Eq. (\ref{Eq:lyapunov}). Shown in
Fig.~\ref{Fig:LyapunovPlot} are solutions of the free energy
$\mathfrak{F} = \int \mathfrak{L} ~ dX dY$
for a polystyrene nanofilm with $h_o = 100 \: \textrm{nm}$ and $\Delta
T = 46 ^\textrm{o}\!\textrm{C}$ for two different wafer separation distances, $d_o =
285$ nm  and 800 nm. The termination points represent $\tau_\textrm{top}$.
The individual contributions to the total free energy (denoted by ``Sum'') from capillary
and thermocapillary terms feature several important points.

For $\tau \lesssim 1$, the film experiences small deformations such that the
opposing capillary and thermocapillary contributions are also small,
neither significantly enhancing nor depleting energy from the evolving film.
Magnified views of the curves (not shown) confirm a small
but monotically decreasing value of the free energy due to the still dominant influence
of thermocapillary stresses. This period of growth
corresponds to the linear regime described by linear stability analysis.
Strong departure from this behavior occurs for $\tau \gtrsim 1$ when nonlinear
effects begin to dominate. In this regime, the time (or distance)
remaining for fluid contact with the top wafer is small and the
energetics of pillar formation strongly affected by the presence of the cooler target.
For the smaller gap separation distance ($d_o$
= 285 nm) shown Fig.\ref{Fig:LyapunovPlot} (a), thermocapillary effects dominate
capillary effects as the nanopillars grow ever more rapidly toward the cooler
target. There remains sufficient fluid in the residual
film to continue feeding the growth of nanopillars such that the system continuously lowers
its overall free energy by transporting fluid toward the cooler
substrate. Unlike the equilibrium cellular convective patterns observed with
Rayleigh-B\'{e}nard or B\'{e}nard-Marangoni instabilities, this nanofilm
instability is non-saturating and the free energy continues to decrease until
the fluid makes contact with the cooler target.

The results shown in Fig. \ref{Fig:LyapunovPlot}(b) for the larger gap separation
distance $d_o = 800$ nm reveal different behavior. Since the top substrate is positioned further away,
the initial thermal gradient is smaller and the films require correspondingly
longer times to develop substantial fluid elongations. The linear to nonlinear
transition is observed to occur at slightly later times, $\tau \approx 1.2$.
The individual contributions to the free energy are still clearly distinguishable
but eventually asymptote. The larger wafer separation distance allows for longer growth
periods, which causes significant film depletion near the base of nanopillars.
Fluid transfer needed to grow the
elongations is impeded, eventually halting their growth.
Fluid already contained within the nanopillars continues to undergo a
circulatory flow pattern, rising upwards near the surface due to thermocapillary
stresses and falling downwards near the interior due to capillary stresses. However,
fluid transfer from the initial deposited film slows considerably and can
be halted completely if the depletion effect causes dryout.

In summary, the Lyapunov analysis demonstrates why there is no steady state configuration
in nanofilms except in cases where film depletion leads to
pillar isolation. This limit can be achieved by placing the secondary plate
sufficiently far from the initial deposited film. In this case, nanopillars
that form will continue to undergo surface and interior flow but they cannot
grow substantially in height due to a limitation in the available fluid mass
needed to feed continued growth and elongation.

\subsection{Influence of relative gap spacing and substrate tilt on symmetry of evolving films}

\subsubsection{Effect of larger gap spacing}

It is interesting to explore further the nonlinear behavior shown in
Fig.~\ref{Fig:growthrate} for times $\tau \gtrsim 1$ by examining images
of the evolved films. The nonlinear regime is characterized by film
deformations that are no longer merely a linear superposition of contributions
with independent wave number.
Instead, the growth of individual peaks influences the growth of neighboring
peaks as determined from Eq.~(\ref{Eq:TCinterface}). The evolving pillars can, for example, reposition
themselves along directions that are energetically favorable in order to
maximize the heat flux through the air/liquid bilayer and in so doing,
can influence the in-plane symmetry. This regime can be investigated by holding all
remaining parameters fixed while increasing $D_o$ so as
to allow the fluid elongations more time to grow before contacting the cooler
substrate. This is easily achieved in the simulations by either increasing the
actual plate separation distance, $d_o$, or reducing the initial film
thickness, $h_o$.

\begin{figure}[htp]
\includegraphics[width=8.5cm]{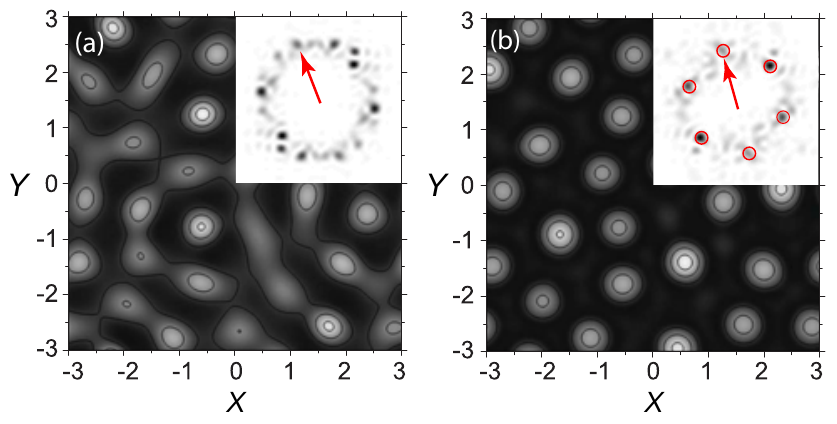}
\caption{Numerical solutions (top view) of the film thickness,
$H(X,Y,\tau_\textrm{top})$, from Eq.~(\ref{Eq:TCinterface})
for different gap ratios, $D_o=d_o/h_o$ for $\Delta
T\!=\!46\:^\textrm{o}\textrm{C}$: (a) $h_o$ = 100 nm, $D_o$ = 3.45,
$\tau_{\textrm{top}}$ = 1.30 and (b) $h_o$ = 40 nm, $D_o$ = 7.125,
$\tau_{\textrm{top}}$ = 1.84. Inset images represent 2D Fourier transforms of
corresponding film heights viewed from above. Fourier coefficients were
normalized to the maximum value for each image and squared for improved
filtering. Unit arrows represent magnitude of most unstable wave
number, $K^\textrm{TC}_\textrm{max}/\sqrt{2} = 2 \pi$, derived from linear
stability theory [see discussion following Eq.
(\ref{Eq:growthrate_simplified})].}
\label{Fig:hexpattern}
\end{figure}

Shown in Fig.~\ref{Fig:hexpattern}(a) and (b) are two representations of the
film height $H(X,Y,\tau = \tau_{\textrm{top}})$ for $\Delta T =
46\:^\textrm{o}\textrm{C}$ and $D_o =
\textrm{(a)}\:3.45\:\textrm{and}\:\textrm{(b)}\:7.125$. The inset figures
depict the corresponding FFTs, where the Fourier coefficients
have been normalized to their peak value and squared for
filtering purposes. The arrow shown has unit length and represents the value
$K^{\textrm{TC}}_{\textrm{max}}$. Contact with the cooler plate is achieved at
$\tau_{\textrm{top}} \approx 1.30\:\textrm{and}\:1.84$,
respectively. The Fourier transform of the pattern (inset) for the smaller
value of $D_o$ suggests  quasi-hexagonal symmetry, with some pronounced
harmonics in the vicinity of the dominant peaks.
By contrast, the pattern for
the larger value of $D_o$ clearly shows well developed hexagonal symmetry.
These patterns indicate that the formation of hexagonal
symmetry is correlated with film depletion near the base of nanopillars.
For some parameter sets investigated,
there is also evidence of a bifurcation cascade, in which the region halfway in
between two adjacent nanopillars generates a parasitic protrusion smaller
in amplitude but similar in shape to the primary nanopillars. This cascade
behavior resembles the dynamics reported in other thin film instabilities
\cite{JooBankoff:JFM1991,KrishnamoorthyRamaswamy:PoF1995,BoosThess:PoF1999}.
For this cascade to occur, the value of $D_o$ must be sufficiently
large such that the growth of the dominant nanopillars consumes a substantial
portion of the interstitial fluid mass.

In reviewing images of nanopillar formation in the literature, it is evident
that hexagonal symmetry can occur with even small values of $D_o$, as shown in
Fig.~\ref{Fig:setup}(d) for which $D_o=1.63$. If the pillars are allowed to
grow well beyond the time required for initial contact with the
cooler substrate, then the dynamics of growth by thermocapillary stresses will
likely continue to draw liquid upwards, thereby thickening the diameter of
nanopillars which bridge the gap in between the two substrates. This process
will continue to remove film material from the interstitial regions thereby
generating conditions favorable to the formation of hexagonal symmetry. In such
cases, the hexagonal symmetry is likely established well after the fastest
growing peaks make contact with the cooler plate. The mechanism leading to this
scenario, however, is not included in the model leading to Eq.~(\ref{Eq:interface}).

\subsubsection{Effect of substrate tilt}

\begin{figure}[htp]
\includegraphics[width=8.5cm]{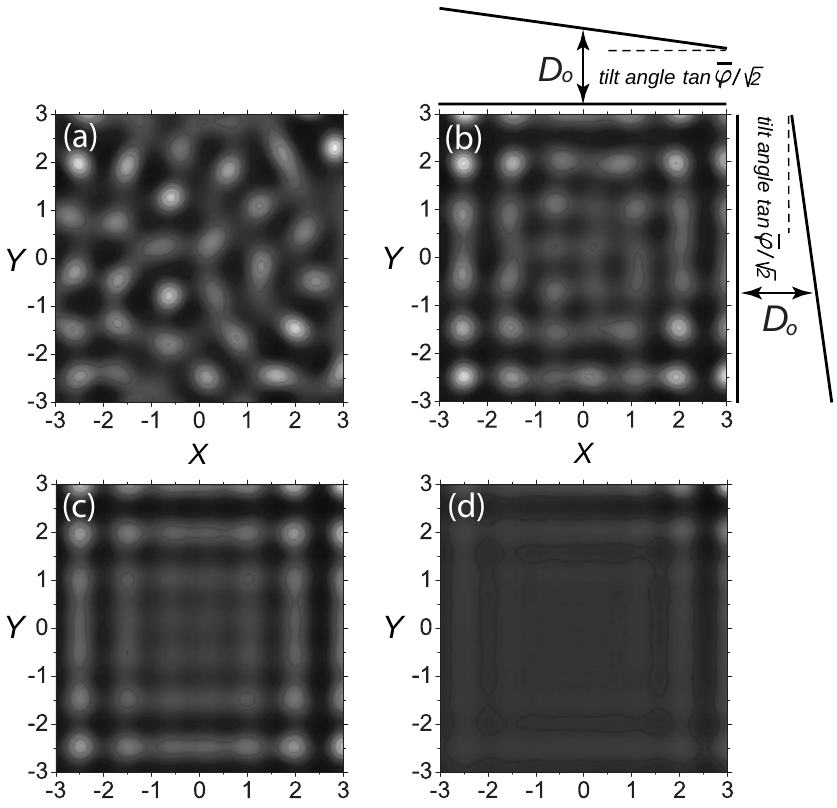}
\caption{Numerical solutions (top view) of the film thickness,
$H(X,Y,\tau_\textrm{top})$, from Eq.~(\ref{Eq:TCinterface_tilt}) for
different inclination angles $\tan(\overline{\varphi})$ of the cooler substrate:
(a) $\tan(\overline{\varphi})\!=\!4.8\cdot10^{-5}$ and
$\tau_\textrm{top}=1.23$, (b) $\tan(\overline{\varphi})\!=\!4.8\cdot10^{-4}$
and $\tau_\textrm{top}=1.12$, (c)
$\tan(\overline{\varphi})\!=\!4.8\cdot10^{-3}$ and $\tau_\textrm{top}=0.86$,
(d) $\tan(\overline{\varphi})\!=\!4.8\cdot10^{-2}$ and
$\tau_\textrm{top}=0.47$. In all cases, $h_o\!=\!100\:$nm, $d_o\!=\!285\:$nm
and $\Delta T\!=\!46\:^\textrm{o}\textrm{C}$. Inclination angle was imposed along
the diagonal of the computational domain such that
$\tan(\overline{\varphi})/\sqrt{2}=\mid\!\partial D/\partial
X\!\mid=\mid\!\partial D/\partial Y\!\mid$. Schematic diagram indicates that upper right corner (lower left
corner) is subject to the smallest (largest) wafer separation
distance.} \label{Fig:tfheight2D_tilt}
\end{figure}

As discussed in Section II.B, the evolution equation for the film height is
modified according to Eq.~(\ref{Eq:TCinterface_tilt}) when the confining
substrates are subject to a relative tilt. Shown in Fig.~(\ref{Fig:tfheight2D_tilt}) are
the corresponding results for solutions of $H(X,Y, \tau_\textrm{top})$ for the
case $h_o = 100$ nm , $d_o = 285$ nm and $\Delta T = 46\:^\textrm{o}\textrm{C}$
subject to increasing inclination angle. The image shown in Fig.~(\ref{Fig:tfheight2D_tilt})(c)
corresponds to the inclination angle used in the experiments of
Sch\"affer \etal
As described in Section II.B, the tilt of
the upper substrate was defined by the unit vector $\overrightarrow{T}_\parallel
= (1,1)/\sqrt{2}$. The top right corner in the images shown corresponds to the region
of the film with the smallest gap separation distance; likewise, the bottom left corner
represents the region with the largest gap distance. As such, a lateral thermal gradient
is established which draws fluid from the bottom left corner into the upper right corner.
To conserve mass in the simulations, fluid exiting the top (bottom) boundary
was simultaneously replaced by fluid entering the right (left) boundary.

In all the experiments of Sch\"affer \etal, the confining substrates were subject
to a relative tilt $\tan(\varphi)
\approx 1\:\mu \textrm{m}/\textrm{cm}$. In
rescaled units, $\tan(\overline{\varphi}) = \tan(\varphi)/\epsilon$ where
$\epsilon = h_o/\lambda^\textrm{TC}_\textrm{max}$. For the experiments in which
$h_o = 100$ nm, $d_o = 285$ nm and $\Delta T = 46\:^\textrm{o}\textrm{C}$,
$\lambda^\textrm{TC}_\textrm{max} = 4.8 ~\mu\textrm{m}$ [Eq.
(\ref{Eq:TCwavelength})], such that $\tan(\overline{\varphi}) = 4.8
\cdot10^{-3}$. The corresponding tilt angle along the $X$ and $Y$ axes for such experiments
corresponds to a value of $4.8 \cdot 10^{-3}/\sqrt{2}\approx 3.3 \cdot 10^{-3}$, which should lead
to the formations observed in Fig.~(\ref{Fig:tfheight2D_tilt})(c) if there were no other
considerations or artifacts.

As evident from the images (b) - (d), the symmetry of the evolving instability
transitions from hexagonal-like to square-like symmetry due to the lateral bias
in thermal gradient established by the tilt of the cooler substrate. Even for
very small tilt angles, fluid is preferentially transported toward the upper
right corner where it accumulates in the form of ridges along the top and right
boundaries. This accumulation process establishes secondary and tertiary parallel ridges
spaced apart roughly by a distance $\lambda^\textrm{TC}_\textrm{max}$.
At longer times, these ridges are observed to undergo breakup with
a similar lateral spacing. A relative tilt of the substrates therefore
introduces a strong lateral bias in thermal gradient which triggers pattern
formation along the domain boundaries instead of within the interior, where the
instability is generally more homogeneously distributed. This specific square
symmetry observed is therefore a direct consequence of using inclined substrates
within a square computational domain.
Modification of the computational domain shape may alter the
symmetry observed; however, the nanopillars will still nucleate along cooler regions
of the film. Additional
studies of the Fourier transforms of the images shown in
Fig.~(\ref{Fig:tfheight2D_tilt})(a) - (d) (not shown) confirm that the fastest
growing wavelength, $\lambda^{\textrm{TC}}_{\textrm{max}}$, remains unaffected
by very small tilt angles. In this respect, the measurements of
$\lambda_{\textrm{max}}$ made by Sch\"affer \etal with a tilted wafer geometry
should not have affected comparison to analytic predictions from linear
stability theory for films confined by parallel wafers. Given the strong
influence of edge behavior on the formation of emerging patterns, however, care
should be taken in experiment to ensure that no artifacts, anomalies or
asymmetries exist along the edges of a film undergoing nanopillar formation if
a particular array symmetry is desirable.

\section{Conclusion}

In this work, we provide evidence that the spontaneous formation of
periodic pillar arrays in molten polymer nanofilms confined within
closely spaced substrates maintained at different temperatures
is due to a thermocapillary instability. If not mass limited, these
pillars continue to grow until contact with the cooler substrate is achieved.
So long as the initial film thickness and substrate separation distance are
sufficiently small that gravitational forces are negligible, there is
no critical number for onset of instability. In contrast with
the conventional B$\acute{e}$nard-Marangoni instability, nanofilms
are prone to formation of elongations no matter
how small the transverse thermal gradient. Ultra small gradients,
however, lead to large values of the most unstable wavelength.
In practice, very large pillar spacings can be difficult to
observe or difficult to distinguish from defect mediated bumps
which also undergo growth from thermocapillary flow.
The linear stability analysis shows that pillar formations are expected
in any viscous Newtonian-like nanofilm. Since the shear rates are characteristically small,
it is expected that molten materials of many kinds can be modeled as a Newtonian fluid.
Pillar arrays formed from polymers like
PS or PMMA are of commercial interest, however, since they solidify rapidly in place once the thermal gradient
is removed due to their lower glass transition temperatures.

The analytic results obtained, including the energetics of nanopillar formation
as described by the Lyapunov functional, confirm
that elongations are caused by the predominance of thermocapillary
stresses, which far outweigh stabilization by capillary stresses
during the later stages of development. The increase in thermocapillary
stresses leads to a rapid decrease in the overall free energy of the
evolving film. Fourier analysis of the emerging structures also indicates
a preference for hexagonal packing although true hexagonal order
cannot be achieved if the separation distance is too small since
the pillars have insufficient time to grow and self-organize before making contact with the
cooler target. Simulations for larger values of $D_o = d_o/h_o$
show well developed and long range hexagonal order. The only limitation of the
current analysis is the restriction to films of constant viscosity. While this
approximation holds well for simple fluids, it is known that the viscosity of
polymer melts like PS and PMMA exhibit a strong dependence on temperature.
It is therefore expected that fluid
elongations undergo an increase in
viscosity as the cooler substrate is approached. We have examined this
effect in detail in a separate study \cite{Dietzel_Visc:JAP2010} and concluded that while this cooling effect
slows the growth of pillars, it does not affect the pillar spacing in any appreciable way.
This is expected since the expression for the
most unstable wavelength given by Eq.~(\ref{Eq:TCwavelength}) is independent of
the melt viscosity.

The linear stability analysis of an initially flat viscous film of thickness
$h_o$ subject only to capillary and thermocapillary forces
reveals that the normalized gap spacing $d_o/h_o$ and temperature drop $\Delta T$
strongly affect the value of the most unstable wavelength,
$\lambda^\textrm{TC}_\textrm{max}$, for given material parameters.
The analysis indicates that the pillar spacing
or array pitch scales as $(\Delta T)^{-1/2}$, which can therefore be tuned in experiment. Direct
comparison of $\lambda^\textrm{TC}_\textrm{max}$ to experimental measurements of Sch\"affer \etal
reveals excellent agreement with the functional dependence on $d_o$, namely
$\lambda^\textrm{TC}_{\textrm{max}}=C_1\:{d_o}^{1/2} + C_2\:{d_o}^{-1/2}$.
The discrepancies observed are attributed to a number of factors including
solvent retention effects in unannealed films and measurements of the array pitch in vitrified
films examined after the fluid had experience prolonged contact with the cooler substrate.
A number of factors not included in the model can influence the
array pitch since the melt is no longer growing in air but migrating and reorganizing along
the underside of the cooled wafer.

A linear stability analysis and numerical solutions of the nonlinear evolution equation were also
conducted for a tilted cooler substrate.
Such a tilt initially establishes both a lateral and vertical thermal gradient.
In the experiments of Sch\"affer \etal, the tilt angle was less than
$0.006 ^o$. Numerical
simulations of the film height for even very small tilt angles confirm
that while the dominant wavelength is unaffected, the in-plane symmetry
of evolving elongations can transition from
hexagonal to square-like symmetry. This change is caused by thermocapillary influx
of fluid into the region subject to a smaller gap width where the effective surface film
temperature is cooler due to closer proximity with the cooler tilted substrate.
The elongations in this region also grow more rapidly since the effective thermal gradient is
larger. These results highlight the
importance of boundary conditions in establishing the in-plane symmetry of arrays formed
as a result of thermocapillary instability in a tilted geometry. This
observation can also be used to advantage to generate large area arrays of
different symmetry.

In conclusion, the results presented here strongly suggest
that thermocapillary stresses play a crucial if not dominant role in the
formation of pillar arrays in molten nanofilms subject to a transverse
thermal gradient. According to the linear stability analysis, nanoscale
films for which the hydrostatic pressure is completely negligible in comparison to
capillary and thermocapillary forces will promote fluid elongations no matter
how small the temperature difference between the top (cooler) and bottom (warmer) substrates.
Experiments using lower viscosity melts, larger thermal graidents,
smaller wafer separation distances, and smaller initial film thicknesses should produce
nanostructures with submicron lateral feature sizes.
We hope that future studies such as these can assist with the
design and fabrication of functional devices by taking advantage of the
inherent regularity, smoothness and robustness of self-organized patterns arising
from a controllable hydrodynamic instability.

\section{Acknowledgement}

The authors gratefully acknowledge financial support for this project from the
CBET Division of the Engineering Directorate of the National Science
Foundation. They also wish to thank the referee for a close reading of this lengthy manuscript.

\renewcommand{\theequation}{A-\arabic{equation}} 
\setcounter{equation}{0}  
\section{Appendix}
To begin, Eq.~(\ref{Eq:TCinterface}) is re-expressed in terms of the parameter
$\chi$ such that
\begin{equation}
\label{Eq:lya_step1} \frac{\partial H}{\partial \tau} = - \nabla_\parallel
\cdot H^3 \left (\frac{\kappa \overline{Ma}}{2 D_o\:H(1 + \chi
H)^2}\nabla_\parallel H +\frac{1}{3 \overline{Ca}} \nabla^3_\parallel H
\right),
\end{equation}
which is rearranged according to
\begin{equation}
\label{Eq:lya_step2}
\begin{split}
\frac{\partial H}{\partial \tau} = -\!\nabla_\parallel\!\cdot\!H^3
\left[\frac{\kappa \overline{Ma}}{2 D_o}\left(\frac{1}{H}\!-
\!\frac{\chi}{1+\chi H}\!-\!\frac{\chi}{(1+\chi H)^2}\right)\nabla_\parallel H \right.\\
\left. + \frac{1}{3 \overline{Ca}}\nabla^3_\parallel H \right].
\end{split}
\end{equation}
The term proportional to $\nabla_\parallel H$ is further simplified, where
\begin{eqnarray}
\label{Eq:lya_step3} \left(\frac{1}{H}\!-\!\frac{\chi}{1+\chi
H}\!-\!\frac{\chi}{(1+\chi H)^2}\right)\nabla_\parallel H = \notag \\
\nabla_\parallel\left[\ln\left(\frac{H}{1+\chi H}\right)\!+\!\frac{1}{1+\chi
H}\right].
\end{eqnarray}
By introducing the function $\Psi=(\kappa
\overline{Ma})/(2D_o)\{\ln[H/(1+\chi H)]+1/(1+\chi H)\}\!+\! C_o$,
the evolution equation can be recast as
\begin{equation}
\label{Eq:lya_step4}
\frac{\partial H}{\partial \tau} = - \nabla_\parallel
\cdot H^3 \left[ \nabla_\parallel \left(\Psi + \frac{1}{3 \overline{Ca}}
\nabla^2_\parallel H\right) \right],
\end{equation}
where $C_o$ is a constant of integration. Eq.~(\ref{Eq:lya_step4})
is then multiplied by the quantity $\tilde{\Psi} = \Psi + \nabla^2_\parallel
H/(3 \overline{Ca})$ to give
\begin{equation}
\label{Eq:lya_step5} \tilde{\Psi} \frac{\partial H}{\partial \tau} = -
\tilde{\Psi}\nabla_\parallel \cdot H^3 \nabla_\parallel \tilde{\Psi}.
\end{equation}
Since $\Psi = \Psi(H)$, one can apply Leibnitz's rule for differentiation to find
\begin{equation}
\label{Eq:lya_step6} \frac{\partial I}{\partial \tau} \equiv
\frac{\partial}{\partial \tau} \int^{H(\tau)}_{H(\tau = 0)} \Psi(S) dS =
\Psi\frac{\partial H}{\partial \tau}
\end{equation}
where $H(\tau = 0)=1$ i.e. the initial film is flat and uniform.
Evaluation of the function $I$ then gives
\begin{eqnarray}
\label{Eq:lya_step7} I \!=\! \frac{\kappa
\overline{Ma}}{2D_o}\left[H\ln\!\left(\frac{H}{1+\chi
H}\!\right)\!+\!\ln(1\!+\!\chi)\right]\!\!+\!C_1(H\!-\!1)
\end{eqnarray}
where $C_1$ denotes a second constant of integration.

Equation (\ref{Eq:lya_step5}) is then integrated over the square domain
$A_\parallel = \Delta X \Delta Y$:
\begin{eqnarray}
\label{Eq:lya_step8} \int_{A_\parallel} \left(\frac{\partial I}{\partial \tau}
+ \frac{1}{3 \overline{Ca}}
\nabla^2_\parallel H \frac{\partial H}{\partial \tau}\right)dX dY = \notag \\
-\int_{A_\parallel} \tilde{\Psi}\nabla_\parallel \cdot H^3 \nabla_\parallel
\tilde{\Psi}dX dY,
\end{eqnarray}
where $\tilde{\Psi}\nabla_\parallel \cdot (H^3 \nabla_\parallel
\tilde{\Psi})$ can be re-expressed as $\nabla_\parallel \cdot
(\tilde{\Psi}H^3\nabla_\parallel \tilde{\Psi}) - H^3(\nabla_\parallel
\tilde{\Psi})^2$. The first term on the right hand side vanishes for a fixed
domain subject to periodic boundary conditions; the integral
$\int_{A_\parallel} (\partial I/\partial \tau) dX dY$ can be
rewritten as $ d/d \tau \int_{A_\parallel} I dX dY$. These simplifications can be
used to recast Eq.~(\ref{Eq:lya_step8}) into
\begin{eqnarray}
\label{Eq:lya_step9}
\frac{d}{d \tau} \int_{A_\parallel} I(H) dX dY + \frac{1}{3 \overline{Ca}}\int_{A_\parallel}
\nabla^2_\parallel H \frac{\partial H}{\partial \tau} dX dY = \notag \\
\int_{A_\parallel} H^3 \left( \nabla_\parallel \tilde{\Psi} \right)^2 dX dY.
\end{eqnarray}
A final integration by parts subject to periodic boundary conditions simplifies the
second integral on the left hand side such that
\begin{eqnarray}
\label{Eq:lya_step10} \int_{A_\parallel}\!\nabla^2_\parallel H \frac{\partial
H}{\partial \tau} dX dY \!\!&=&\!\!\! -\frac{1}{2} \int_{A_\parallel}\,
\!\! \frac{\partial}{\partial \tau}\left(
\nabla_\parallel H \right)^2 dX\!dY \notag \\
\!\!&=& \!\!\!-\frac{1}{2} \frac{d}{d \tau} \int_{A_\parallel}\!\!\!\left( \nabla_\parallel H
\right)^2 dX\!dY
\end{eqnarray}
Equation~(\ref{Eq:lya_step9}) then simplifies to the form
\begin{eqnarray}
\label{Eq:lya_step11} \frac{d}{d \tau} \int_{A_\parallel}
\left[I - \frac{1}{6 \overline{Ca}} \left(\nabla_\parallel H\right)^2\right] dX dY = \notag \\
\int_{A_\parallel} H^3 \left( \nabla_\parallel \tilde{\Psi} \right)^2 dX dY.
\end{eqnarray}
Inserting Eq.~(\ref{Eq:lya_step7}) into Eq. (\ref{Eq:lya_step11}) and noting
that volume conservation within the domain $A_\parallel$ requires that
$\int_{A_\parallel} (H-1)dX dY = 0$ leads to
\begin{equation}
\label{Eq:lya_step12}
\begin{split} \frac{d}{d \tau}
\int_{A_\parallel}
\left\{\frac{\kappa\overline{Ma}}{2D_o}\left[H\ln\left(\frac{H}{1+\chi
H}\right)+\ln(1+\chi)\right] \right. \\
\left. - \frac{1}{6 \overline{Ca}} \left(\nabla_\parallel H\right)^2\right\} dX dY = \\
\int_{A_\parallel} H^3 \left( \nabla_\parallel \tilde{\Psi} \right)^2 dX dY.
\end{split}
\end{equation}
Multiplying Eq. (\ref{Eq:lya_step12}) by the quantity $-6~\overline{Ca}$ produces
the final expression for the rate of change of $\mathfrak{F}$, namely
\begin{equation}
\label{Eq:lya_final_1} \frac{d}{d \tau} \int_{A_\parallel} \mathfrak{L} dX dY =
-6\overline{Ca}\int_{A_\parallel} H^3 \left( \nabla_\parallel \tilde{\Psi}
\right)^2 dX dY \leq 0,
\end{equation}
where $\mathfrak{L}$ is given by Eq. (\ref{Eq:lyapunov}).
Since Eq.~(\ref{Eq:lya_final_1}) is a non-negative quantity,
the thin film seeks configurations of the interface $H$ in time which minimize
$\mathfrak{F}$.


\end{document}